\documentclass[prb, aps,
twocolumn,
longbibliography,
amsmath,amssymb,
floatfix
]{revtex4-2}

\usepackage{physics}
\usepackage{epsf}
\usepackage{float}
\usepackage{graphicx}
\usepackage{amsmath}
\usepackage[usenames]{color}
\usepackage{float}
\usepackage{bm}
\usepackage{relsize}
\usepackage[normalem]{ulem}
\usepackage{mathrsfs}

\def \be {\begin{equation}}
\def \ee {\end{equation}}
\def \bea {\begin{align}}
\def \eea {\end{align}}

\def \BEA {\begin{eqnarray}}
\def \EEA {\end{eqnarray}}
\def \BC {\begin{cases}}
	\def \EC {\end{cases}}

\usepackage{siunitx}
\usepackage{tabularx}
\usepackage{xcolor}
\usepackage[version=4]{mhchem}
\usepackage{ctable}

\usepackage[unicode=true,colorlinks=true,citecolor=blue,urlcolor=blue]{hyperref}

\def\be {\begin{equation}}
\def\ee {\end{equation}}
\def\bea {\begin{align}}
\def\eea {\end{align}}

\def\bee{\begin{eqnarray}}
\def\eee{\end{eqnarray}}
\def\BC {\begin{cases}}
	\def\EC {\end{cases}}

\makeatother

\begin{document}
	\title{Nonlinear helicity anomalies in the cyclotron resonance photoresistance of two-dimensional electron systems
 }

	\author{E.~M{\"o}nch$^1$, S.~Schweiss$^1$, I. Yahniuk$^1$, M.~L.~Savchenko$^2$, I.~A.~Dmitriev$^1$, A.~Shuvaev$^2$, A.~Pimenov$^2$, D.~Schuh$^1$, D.~Bougeard$^1$ and S.~D.~Ganichev$^1$}
	
	\affiliation{$^1$Terahertz Center, University of Regensburg, 93040 Regensburg, Germany}
	\affiliation{$^2$Institute of Solid State Physics, Vienna University of Technology, 1040 Vienna, Austria}
	
	

	\date{\today}
	
	\begin{abstract}	
		Our studies of the cyclotron resonance (CR) photoresistance in GaAs-based two-dimensional electron systems (2DES) reveal an anomalously low sensitivity to the helicity of the incoming circularly polarized terahertz radiation. We find that this anomaly is strongly intensity dependent, and the ratio of the low-temperature photoresistance signals for the CR-active (CRA) and CR-inactive (CRI) polarities of magnetic field increases with lowering power, but, nevertheless, remains substantially lower than expected from conventional theory assuming interaction of the plane electromagnetic wave with the uniform 2DES. Our analysis shows that all data can be well described by the nonlinear CR-enhanced electron gas heating in both CRA and CRI regimes. This description, however, requires a source of anomalous absorption of radiation in the CRI regime. It can stem from evanescent electromagnetic fields originating from the near-field diffraction within or in the vicinity of the quantum well hosting the 2DES.   
	\end{abstract}
	
	\maketitle

	\section{Introduction}

The electron cyclotron resonance (CR) is one of the central phenomena in magnetooptics, in particular, providing a standard method to measure the energy dispersion of carriers in three (3D) and two (2D) dimensions conducting materials, see, e.g., Refs.~\cite{Seeger2004,Hilton2012}. The resonance occurs when the frequency of radiation matches that of the cyclotron motion of carriers in a static magnetic field. Importantly, in the Faraday geometry and for the circularly polarized electromagnetic (EM) wave the CR emerges for one magnetic field polarity only. Specifically, the electrons should perform a helical (3D) or circular (2D) motion with the same sense of rotation as the driving force of the EM wave, as only in this case the electric and magnetic forces continuously match. This is called the CR-active (CRA) regime. Consequently, for the opposite polarity of the magnetic field or opposite helicity of the EM wave the CR should be absent, which is usually called the CR-inactive (CRI) regime.

The main experimental approaches to CR studies include all-optical transmission or reflection spectroscopy, the optically detected cyclotron resonance (ODCR), and photoelectric methods, for review see Ref.~\cite{Hilton2012}. The latter, studying the CR photoresistance or photocurrent, are particularly important for exploration of modern 2D materials. Apart from the energy spectrum, the photoelectric methods provide valuable information on elastic and inelastic scattering processes controlling the momentum and energy relaxation. Moreover, such studies provide access to rich spectrum of CR-related nonequilibrium phenomena such as microwave induced resistance oscillations (MIRO)~\cite{Zudov2001,Dmitriev2012}, magnetoplasmon effects including excitation of non-local Bernstein modes~\cite{Bandurin2022}, non-Markovian classical memory effects~\cite{Dmitriev2008}, to name a few. Stunningly, recent studies of these effects in 2D electron systems (2DES) revealed up to 100\% immunity of the CR response to radiation helicity, first in MIRO studies~\cite{Smet2005,Herrmann2016,Herrmann2017} and, most recently, also in the CR photoresistance and photocurrent~\cite{Moench2022b}. 

In all these experiments, the CR helicity anomaly was detected solely in the photoelectric response, whereas the simultaneous transmittance measurements demonstrated regular strong helicity dependence, in full agreement with the well established Drude theory of the CR. The anomaly was detected even in 2DES samples with lateral size much larger than the beam spot of the radiation, which excludes depolarization effects due to diffraction on sample edges or contacts as a possible origin of the anomaly. 

 Here we report on a detailed study of the CR helicity anomaly. Studying the terahertz radiation-induced change of the low-frequency resistance in high-mobility GaAs/AlGaAs quantum well (QW) structures we observed that the ratio of the CR signals in the CRA and CRI regimes (CRA/CRI ratio) crucially depends on the radiation power, $P$. While at high $P$ and low temperatures, $T$, the CRA/CRI ratio can be close to unity (100\% immunity), it substantially increases at low $P$ and/or high $T$. Importantly, even at the lowest powers the CRA/CRI ratio remains anomalously low.  We demonstrate that the observed power dependence of the CRA/CRI ratio is caused by saturation of the photoresistance which has different saturation powers for the CRA and CRI polarities. 
 
 Our analysis shows that, in the CRA regime, both photoelectric response and the magnetotransmittance can be consistently described using the standard theory, with the nonlinear power dependence caused by uniform electron gas heating by the plane circularly polarized EM wave due to nonlinear energy losses. On the other hand, the observed behavior of the resonant photoresistance in the CRI regime appears to be very similar to that in the CRA regime, but with a smaller overall magnitude and different saturation intensity. Thus, it should be attributed to the same physical mechanisms, but with a weaker CRI resonant absorption in comparison to the CRA regime. Microscopic mechanisms of such anomalous absorption were discussed in Ref.~\cite{Moench2022b} and involve the emergence of significant evanescent-wave components localized near the 2DES plane. Such diffracted near-field components of the wave should have a different polarization with respect to the incoming circularly polarized wave. In agreement with our findings, they enable the observed resonant CR absorption for both polarities of magnetic field and, simultaneously, do not affect the regular behavior of transmitted wave measured in the far field. 
 
 			\begin{figure}
		\centering \includegraphics[width=\linewidth]{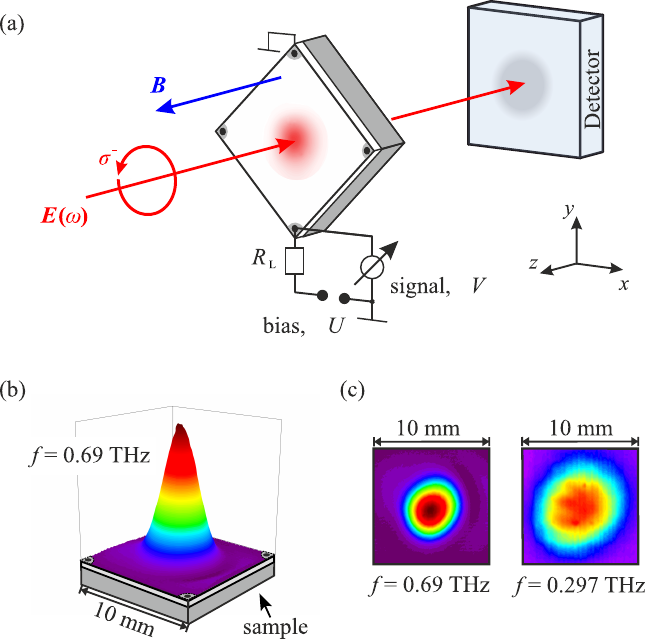}
		\caption{
			(a) Sketch of the experimental setup used for simultaneous measurements of the photoresistance and radiation transmittance. 
  Most experiments were performed in 2-terminal configuration, apart from several measurements in 4-terminal configuration which will be mentioned in text below. (b) Three-dimensional intensity profile of the laser beam at $f = 0.69$~THz, focused at the 10$\times$10~mm$^2$ GaAs sample, measured by the pyroelectric camera. (c) Corresponding beam cross sections recorded for $f = 0.69$ and 0.297~THz yielding respective spot diameters (full width at half maximum) of $d=3.0$ and $5.8$~mm.}
		\label{FigS1_2} 
	\end{figure}
		
	\begin{table*}
		\centering
		\begin{tabular}{|c|c|c|c|c|c|}
			\hline
			Sample       & QW thickness   & with room light&   $n_e$ &$\mu$ & $\tau_p$   \\
			& (nm) & illumination & 10$^{11}$\,(cm$^{-2}$) & 10$^6$\,(cm$^2/$Vs) & (ps) \\
			\hline
			\#A
			C512C
			& 10 & no &7 & 0.37 & 13  \\
			\#A
			C512C
			& 10 & yes &12 & 0.66 & 24  \\
			\#B
			1614
			& 16 & no & 7 & 1 & 36 \\
			\#B
			1614
			& 16 & yes & 7 & 1 & 36 \\
			\hline
		\end{tabular}
		\caption{Electron density $n_e$, mobility $\mu$, and momentum relaxation time $\tau_p$ obtained from magnetotransport data at $T=1.8$~K. Third column indicates whether data were obtained in the dark or after short illumination by room light. Note that in sample \#B such illumination does not lead to essential changes of $n_e$ and $\mu$, but still promotes the visibility of the quantum effects such as SdHO and MIRO.  
		}
		\label{transport_para}
	\end{table*}

	\section{Samples and Methods}

	%
	%

For our study on the CR we used large-sized GaAs-based heterostructures of different quantum well (QW) thicknesses and fabrication. The first (sample \#A) one consists of an AlGaAs/GaAs QW structure with a thickness of $10$~nm, while the second one (sample \#B) is made of a selectively doped 16-nm GaAs QW with AlAs/GaAs superlattice barriers, both grown by molecular-beam epitaxy (MBE) on a GaAs substrate. For more information on samples, see Refs.~\cite{Moench2022b,Savchenko2022}. The samples were prepared in a van der Pauw geometry both exhibiting a size of $10\times10$~mm$^2$ exceeding the diameter of the spot of the incoming THz or sub-THz beam. Ohmic contacts made out of indium and Ge/Au/Ni/Au for the 10-nm and 16-nm GaAs sample, respectively, were applied to the corners. The transport parameters such as electron density and mobility obtained by conventional magnetotransport measurements are collected in Table \ref{transport_para}. The photoresistance and transmittance experiments were performed on samples in the darkness or after short illumination by room light. The latter leads to a change of transport characteristics due to the persistent photoconductivity effect~\cite{Mooney1990}.

As radiation sources we used a tunable continuous wave ($cw$) impact ionization avalanche transit time diode (IMPATT diode) based system~\cite{Shur1990,Sze2008}, an optically pumped $cw$ molecular gas laser~\cite{Chantry1984,Ganichev2005}, and backward-wave oscillators (BWOs)~\cite{Bruendermann2012}. The former consists of a local oscillator operating at $7.3$ to $14.6$~GHz and a frequency multiplier chain increasing the frequency by 24 times. The system yields continuously tunable radiation operating in the range from $f = 0.28$ to $0.312$~THz and having a maximum power up to $P = 30$~mW. The molecular laser operates at a frequency of $f=0.69$~THz providing powers up to $P = 8$~mW. 
The BWO system provided radiation with $f = 0.35$~THz and a power of $P\sim 0.5$~mW at the sample surface~\cite{Savchenko2022}. 
	
The CR experiments were performed in a temperature-regulated Oxford Cryomag optical cryostat equipped with $z$-cut crystal quartz windows, which were covered by black polyethylene films. The films are almost transparent in the THz range, but block uncontrolled illumination by the ambient light. Photoresistance and transmittance measurements were performed in the Faraday geometry with a magnetic field $B$ up to 7~T applied perpendicularly to the QW plane, see Fig.~\ref{FigS1_2}(a). The sample was illuminated from the top side containing the 2DES by monochromatic THz radiation with frequencies in the range from $f = 0.28$ to 0.69~THz and radiation powers up to $P = 30$~mW. The radiation was focused on the sample by an off-axis parabolic mirror. The intensity distribution and the spot size of the beam were monitored by a pyroelectric camera. Depending on the radiation frequency, the spot diameter varied between $d=3$ and 5.8~mm. For all frequencies it was substantially smaller than the lateral dimensions of the square-shaped samples ($10\times10$~mm$^2$), see Fig.~\ref{FigS1_2}(b). This allows us to exclude possible contributions from edge and contact illumination in photoresistance and transmittance measurements. The transmitted radiation was detected by a pyroelectric detector placed behind the sample, simultaneously with the photoresistance measured in 2-point configuration, see Fig.~\ref{FigS1_2}(a).
	
In the experiments in Regensburg described below we used left- ($\sigma^-$) and right- ($\sigma^+$) handed circular polarized radiation. To modify the initial linear polarization state, a waveguide polarizer was directly connected to the radiation output of the IMPATT diode-based system, while the molecular laser setup utilized $x$-cut crystal quartz quarter wave plates.

   Experiments performed in Vienna employed coherent BWO radiation and an optical cryostat with mylar windows covered by black polypropylene plates. A polarization transformer, consisting of a tunable plane mirror placed in parallel to a fixed wire grid polarizer, was used to create the circular polarization~\cite{Smet2005, Savchenko2022}. The samples were irradiated from the substrate side. The photo-transport measurements were  performed using the double modulation technique and 4-terminal configuration,  see \cite{Kozlov2011, Otteneder2018} and Supplemental Material in Ref.~\cite{Savchenko2021}. 
 Transport measurements performed in Vienna and Regensburg in the absence of THz illumination confirmed that the samples parameters remained the same during measurements in both locations.
 
 To vary the radiation power we used either an internal waveguide attentuator (IMPATT) or a so-called cross-polarizer technique (molecular laser, BWO). The latter consists of two wire grid polarizers, where the rotation of the first one modifies the radiation power, while the second one was fixed to ensure an unchanged output polarization~\cite{Hubmann2019, Candussio2021a}.
	
		\begin{figure*}
		\centering \includegraphics[width=0.75\linewidth]{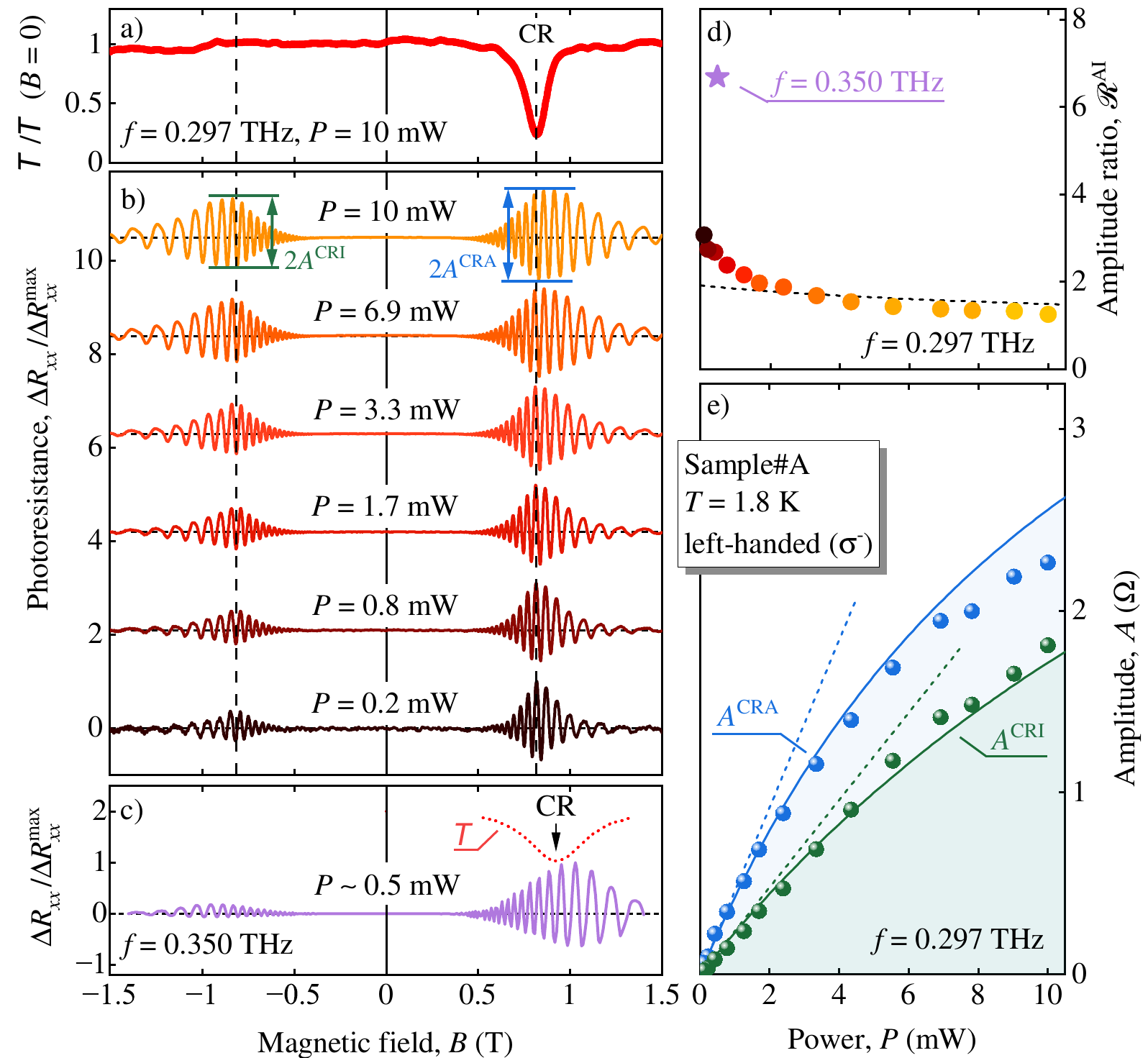}
		\caption{Results obtained in sample \#A at $T=1.8$~K without prior room light illumination. Panels (a) and (b) show the data measured  under
  left-handed circularly polarized radiation ($\sigma^-$) produced by the IMPATT diode operating at a frequency $f = 0.297$~THz. 
  (a) Radiation transmittance recorded at radiation power of $P = 10$~mW and normalized to it's value at zero magnetic field, $\mathcal{T}/\mathcal{T}(B=0)$. 
  (b) Photoresistance measured at various radiation powers. The traces are normalized to the signal's maximum, $\Delta R_{xx}/\Delta R_{xx}^{\text{max}}$ and are up-shifted for clarity. The vertical dashed line on the right side indicates the position of the CR obtained from the minimum of the transmittance. The left vertical dashed line marks the corresponding CR position for the opposite helicity. The blue and green vertical double arrows illustrate how the amplitudes, $A^{\text{CRA}}$ and $A^{\text{CRI}}$, of the resonant photoresistance signals are determined on the CRA and CRI sides. 
  (c) Photoresistance (purple trace) obtained using BWO operating at $f = 0.350$~THz and $P\simeq 0.5$~mW. The red dotted line shows the corresponding CR dip in the simultaneously measured transmittance. 
  (d)  Power dependence of the ratio $\mathscr{R}^{\rm AI}= A^{\text{CRA}}/A^{\text{CRI}}$ of the CRA and CRI amplitudes. The circles refer to the data measured at $f = 0.297$~THz [selected traces are shown in panel (b)], whereas the purple star represents the photoresistance measured at $f = 0.350$~THz [panel (c)]. 
  (e) Power dependence of $A^{\text{CRA}}$ (blue circles) and $A^{\text{CRI}}$ (green circles), as extracted from unnormalized photoresistance data $\Delta R_{xx}$ measured at $f = 0.297$~THz.  Solid lines are calculated according to Eq.~\eqref{saturation} using $a$ and $P_s$ as fitting parameters ($a^\text{CRA} = 0.46$~$\Omega$/mW and $P_s^\text{CRA} = 12.5$~mW at the CRA side,  $a^\text{CRI} = 0.24$~$\Omega$/mW and $P_s^\text{CRI} = 25$~mW at the CRI side). Dashed lines illustrate the linear part $A=a P$ of the corresponding fits. The power dependence of the ratio $A^{\text{CRA}}/A^{\text{CRI}}$ of the fits is illustrated by a dashed curve in panel (d). 	
				}
		\label{FigR1} 
	\end{figure*}

	\begin{figure*}
		\centering \includegraphics[width=0.8\linewidth]{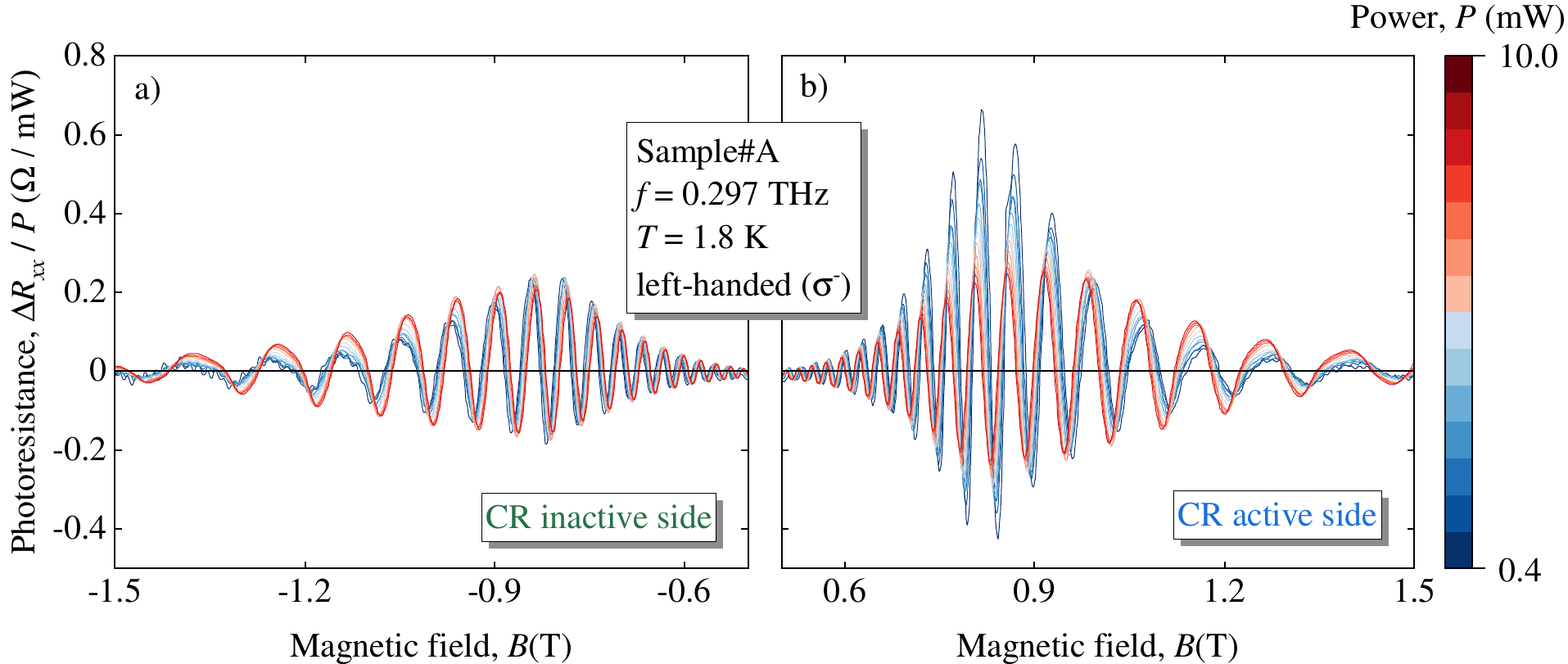}
		\caption{Power dependence of the photoresistance $\Delta R_{xx}$ normalized to the radiation power $P$. Here the same data set as in Fig.~\ref{FigR1}(b) is used.  Panels (a) and (b) show the resonant photoresponse for CRI and CRA configurations, respectively. The color of individual traces changes from intense blue measured at the lowest ($P=0.4$~mW) to intense red at the highest power ($P=10$~mW), as illustrated by the color bar on the right. 
		}
		\label{FigR13} 
	\end{figure*}
	
	\begin{figure*}
		\centering \includegraphics[width=0.8\linewidth]{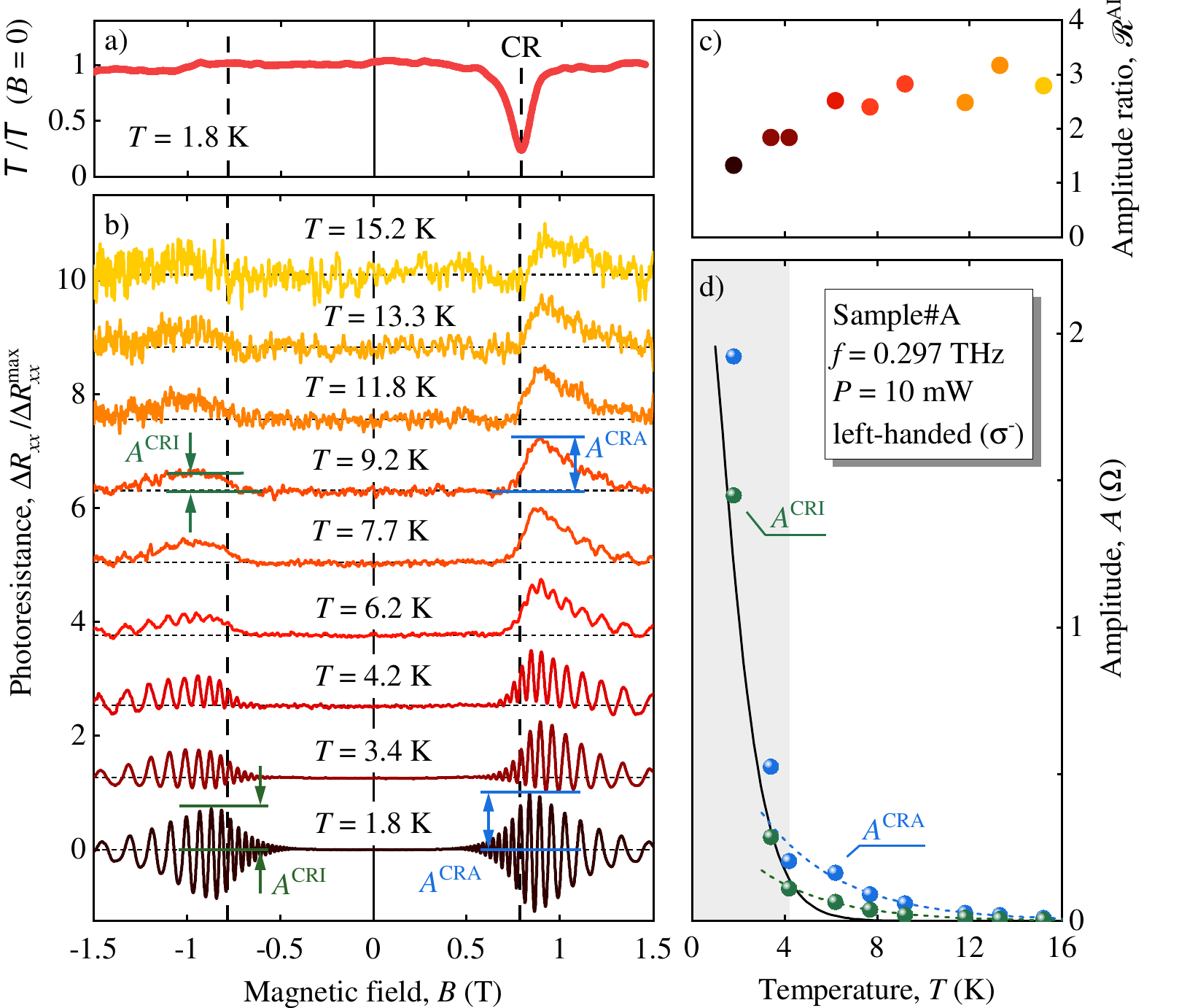}
		\caption{Results obtained in sample \#A (without prior exposure to room light) under left-handed circularly polarized radiation ($\sigma^-$) produced by the IMPATT diode operating at frequency $f = 0.297$~THz with $P=10$~mW. (a) Radiation transmittance normalized to it's value at zero magnetic field, $\mathcal{T}/\mathcal{T}(B=0)$ [same as in Fig.~\ref{FigR1}(a)]. (b) Normalized photoresistance traces, $\Delta R_{xx}/\Delta R_{xx}^{\rm max}$, at given temperatures from $T=1.8$~K up to $15.2$~K. Vertical dashed lines mark the position of the CR for both helicities. Blue and green vertical double arrows (see traces for $T=1.8$~K and 9.2~K) show how the amplitudes $A^{\text{CRA}}$ and $A^{\text{CRI}}$ are determined from the unnormalized data for $\Delta R_{xx}$. (c) Temperature dependence of the amplitude ratio $\mathscr{R}^{\rm AI}= A^\text{CRA}/A^\text{CRI}$. (d) Temperature dependence of individual amplitudes $A^{\text{CRA}}$ and $A^{\text{CRI}}$. The grey shaded area marks the interval of $T \leq 4.2$~K where the SdHO-periodic oscillations in $\Delta R_{xx}$ remain strong. The corresponding solid line is calculated using the Lifshitz-Kosevich formula, see Sec.~\ref{discussion}.
        The dashed blue and green curves serve as guide for the eye, in the region where the SdHO are fully suppressed. 
					}
		\label{FigR11} 
	\end{figure*}

	\begin{figure*}[t]
		\centering \includegraphics[width=0.8\linewidth]{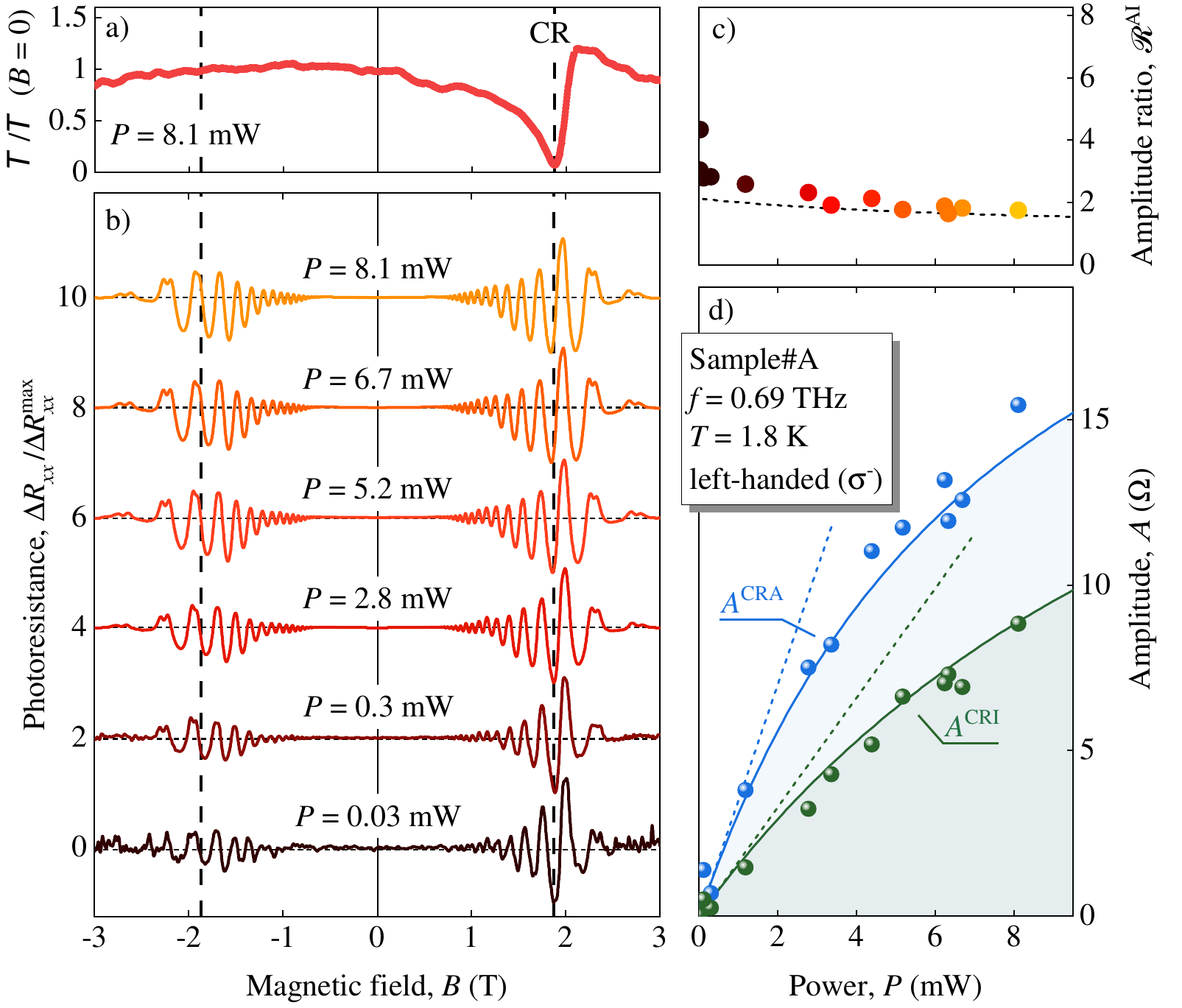}
		\caption{Results obtained in Sample \#A at $T=1.8$~K without exposure to room light. The data are obtained using left-handed circularly polarized ($\sigma^-$) radiation with frequency $f = 0.69$~THz produced by the molecular laser. (a) Normalized transmittance obtained for the highest available power $P = 8.1$~mW. (b) Photoresistance traces measured at various radiation powers. The traces are normalized on the signal's maximum, $\Delta R_{xx}/\Delta R_{xx}^{\text{max}}$, and are up-shifted for clarity. Vertical dashed lines mark the position of the CR for both helicities. (c) Power dependence of the amplitude ratio, $\mathscr{R}^{\rm AI} = A^{\text{CRA}}/A^{\text{CRI}}$. The amplitudes are obtained as illustrated in Fig.~\ref{FigR1}. (d) Power dependence of $A^{\text{CRA}}$ (blue circles) and $A^{\text{CRI}}$ (green circles) extracted from unnormalized photoresistance data. Solid lines are calculated according to Eq.~\eqref{saturation} using $a$ and $P_s$ as fitting parameters ($a^\text{CRA} = 3.51$~$\Omega$/mW and $P_s^\text{CRA} = 8.0$~mW at the CRA side, $a^\text{CRI} = 1.64$~$\Omega$/mW and $P_s^\text{CRI} = 16.0$~mW at the CRI side). Dashed lines illustrate the linear part $A=a P$ of the corresponding fits. The power dependence of the ratio $A^{\text{CRA}}/A^{\text{CRI}}$ of the fits is illustrated by a dashed curve in panel (c).
		}
		\label{FigR5} 
	\end{figure*}

	\section{Results}

Below we present the results of our measurements. We start with the THz photoresistance data obtained in sample~\#A without exposure to room light, Sec.~\ref{sample_A}. Under this conditions the photoresistance is mainly caused by electron gas heating due to radiation absorption at the CR (bolometric response). Next, we describe the results obtained for the same sample, but exposed to ambient light for a short time, Sec.~\ref{sample_A_room}. The latter changes the transport characteristics and gives rise to MIRO in the photoresistance. In the same section we present the results obtained in sample~\#B after brief illumination by room light.

	\subsection{THz response of sample \#A without prior exposure to room light}
	\label{sample_A}
	
Figure~\ref{FigR1}(a) shows the normalized transmitted signal, $\mathcal{T}/\mathcal{T}(B=0)$, recorded at $T = 1.8$~K for sample \#A. It was obtained using the IMPATT diode system operating at a frequency of $f = 0.297$~THz, at the highest available power of $P = 10$~mW and for left-handed circularly polarized radiation ($\sigma^-$). The CR (centered at $B=B_\text{CR}$, indicated by a vertical dashed line) is clearly visible as a dip at the CRA side, which corresponds to positive $B$ for $\sigma^-$ polarization. As discussed in more detail in Sec.~\ref{discussion}, the shape of the measured transmittance can be well reproduced by standard Drude theory incorporating the effects of strong metallic reflection from high-mobility and high-density 2DES as well as multiple reflections of the plane EM wave within the dielectric substrate (Fabry-P\'{e}rot interference)~\cite{Abstreiter1976, Herrmann2016}. In this particular case, the CR in the transmittance has a relatively narrow width that is caused by destructive interference in the substrate. Importantly, the transmittance is almost flat at the CRI side, i.e., at $B<0$, including the vicinity of $B=-B_\text{CR}$ (dashed line). This ensures a high purity of the $\sigma^-$ polarization state of the transmitted wave and, therefore, also of the incoming wave on the opposite side of the sample. 
		
Figures~\ref{FigR1}(b) and \ref{FigR11}(b) display traces of the photoresistance, $\Delta R_{xx}$, 
obtained via varying the radiation power $P$ at fixed temperature $T=1.8$ K and for various $T$ at fixed $P=10$ mW, correspondingly. The traces are offset for clarity and normalized to the maximum positive signal on the CRA side to facilitate the comparison. First of all, it is immediately seen that, in sharp contrast to the simultaneously measured transmittance, Figs.~\ref{FigR1}(a) and \ref{FigR11}(a), the resonant response in $\Delta R_{xx}$ is detected for both CRA and CRI magnetic field polarities and, quite surprisingly, has similar magnitudes. A quick inspection shows that the shape of the resonant photoresistance appears to be very similar on both sides, whereas the relative magnitude of the CRA and CRI signals is highly sensitive to temperature and radiation power. 

On the CRA side, the behavior of $\Delta R_{xx}$ is rather ordinary and can be directly attributed to the conventional resonant CR absorption and associated heating of the electron gas by circularly polarized radiation (bolometric effect), see, e.g., Ref.~\cite{Ganichev2005} and Sec.~\ref{discussion} for details. At high $T$ the photoresistance exhibits a resonant peak, see several upper traces in Fig.~\ref{FigR11}(b), that is associated with the heating-induced decrease of the mobility of 2DES ($\mu$-photoconductivity \cite{Ganichev2005}). At low $T$, Figs.~\ref{FigR1}(b,c) and lower traces in Fig.~\ref{FigR11}(b), the electron gas heating primarily leads to resonant suppression of Shubnikov-de Haas oscillations (SdHO), resulting in the emergence of SdHO-periodic sign-alternating oscillations in $\Delta R_{xx}$. Similar to the resonant peak at high $T$, the envelope of oscillations at low $T$ reflects the $B$-dependence of radiation-induced changes of the electron temperature.  At intermediate  temperatures the photoresponse is caused by the superposition of the $\mu$-photoconductivity and the suppresion of SdHO.

For quantitative analysis of the anomalous relative magnitude of the CRA and CRI signals in Figs.~\ref{FigR1}(b) and \ref{FigR11}(b), we define their amplitudes $A^{\rm CRA}$ and $A^{\rm CRI}$ as the peak height at high $T$ and the maximum value of the envelope of the SdHO-related oscillations at low $T$ and introduce their ratio
	\begin{equation}
	\mathscr{R}^{\rm AI} = \frac{A^{\rm CRA}}{A^{\rm CRI}}\,.
 \label{ratio}
\end{equation}
Figures~\ref{FigR1}(d) and \ref{FigR11}(c) reveal that this ratio depends on the radiation power and on the sample temperature, approaching unity at the highest power and lowest temperature. 

We first address the power dependence. It is seen that the ratio substantially increases with lowering the power. Such a behavior indicates that both amplitudes depend nonlinearly but in a distinct way on the radiation power. Indeed, plots of individual amplitudes $A^{\rm CRA}$ and $A^{\rm CRI}$ in Fig.~\ref{FigR1}(e) show that the CRA response substantially saturates with $P$ while for the CRI response the nonlinearity is weaker. At high power, the observed nonlinearity on both sides can be well fitted using the empirical formula
\begin{equation}
	A(P) = \frac{a P}{1+P/P_s}\,,
	\label{saturation}
\end{equation}
with two fitting parameters $a$ and $P_s$ on each side describing the linear behavior at low $P$ and the saturation power, correspondingly. Blue and green curves in Fig.~\ref{FigR1}(e) show the corresponding fits for $A^{\rm CRA}$ and $A^{\rm CRI}$. In agreement with the discussion above, these fits show that the saturation power on the CR-active side, $P^{\rm CRA}_s = 12.5$~mW, is two times smaller than that on the CRI side, $P^{\rm CRI}_s = 25$~mW. At the same time, the linear coefficients in these fits are $a^{\rm CRA}= 0.46 $~$\Omega$/mW and $a^{\rm CRI}= 0.24$~$\Omega$/mW, so that their ratios yield $P^{\rm CRA}_s/P^{\rm CRI}_s\simeq a^{\rm CRI}/a^{\rm CRA}$. It worth mentioning that, while Eq.~\eqref{saturation} describes well the nonlinear behavior at high power, at low powers it fails, and the linear regime is not reached even at the lowest power used in our experiments. Correspondingly, the CRA/CRI ratio in Fig.~\ref{FigR1}(d) does not saturate completely at low $P$ as it should in the linear regime. 


As mentioned before, at first glance the shape of the resonant photoresistance for CRA- and CRI-configurations appears to be very similar. For a closer inspection of this shape and its evolution with $P$, in Fig.~\ref{FigR13}(a, b) we replotted the photoresistance data from Fig.~\ref{FigR1}(b) without the offset, and with each trace normalized to the corresponding radiation power $P$. This procedure reveals that the nonlinearity of $\Delta R_{xx}$ is pronounced only in the vicinity of $B=B_\text{CR}$ in the CRA configuration, Fig.~\ref{FigR13}(b). It is almost absent in the CRI configuration, as well as at the wings of the CRA signal: here all traces of $\Delta R_{xx} /P$ almost coincide. These observations are in line with the discussion above. Namely, on the CRA side, the electron gas heating is most pronounced at the maximum of CR absorption. At the wings the absorbed power substantially drops and, consequently, the nonlinearity does not show up. At the CRI side, the nonlinearity is weak even at the maximum, see Fig.~\ref{FigR1}(e), so the normalized signals $\Delta R_{xx} /P$ behave almost identically at low and high power.

In addition to the data obtained at $f = 0.297$~THz, we performed measurements using the BWO setup providing $\sigma^-$ radiation at $f = 0.350$~THz and $P \approx 0.5$~mW, see Fig.~\ref{FigR1}(c). In transmittance, the CR dip is also present in the CRA configuration only. Comparing the shape of the dip with that obtained at $f = 0.297$~THz, see Figs.~\ref{FigR1}(c) and (a), we see that at $f = 0.350$~THz the CR is substantially broader indicating the constructive Fabry-P\'{e}rot interference for this frequency, see Sec.~\ref{discussion}.
The observed ratio, $\mathscr{R}^{\rm AI} = 6.5$, in this case was larger than that obtained for $f = 0.297$~THz, see Fig.~\ref{FigR1}(d). 

We now turn to the temperature dependence of the ratio $\mathscr{R}^{\rm AI}$, see also Ref.~\cite{Moench2022b} as well as Figs.~\ref{FigR12} and \ref{FigR13_3} in the Appendix. Figure~\ref{FigR11}(c) obtained at $P = 10$~mW shows that it increases with $T$. The temperature dependences of the individual amplitudes, $A^{\rm CRA}$ and $A^{\rm CRI}$, are shown in Fig.~\ref{FigR11}(d). They depict that the magnitude of the photoresistance signals on both sides rapidly decreases with growing $T$. A typical example of the evolution of the shape of the resonant photoresistance with power at higher temperature ($T=6$~K) is provided in the Appendix, see Fig.~\ref{FigR13bis} there. 

The results discussed so far were obtained primarily by applying radiation at $f = 0.297$~THz. 
Figure \ref{FigR5} demonstrates that the photoresistance in response to radiation of higher frequency, $f = 0.69$~THz, shows a similar behavior. In particular, the ratio $\mathscr{R}^{\rm AI}$ also increases with the reduction of the radiation power by about 2.5 times, see Fig.~\ref{FigR5}(c). The maximum power at both frequencies was almost the same. However, the spot diameter, $d=3$~mm, at $f = 0.69$~THz was substantially smaller than at $f = 0.297$~THz, where $d=5.8$~mm. Thus, despite the radiation power is similar in both cases, the intensity of $f = 0.69$~THz illumination is about 4 times higher. While the saturation powers at the CRA side, obtained using the fits according to Eq.~\eqref{saturation}, are also found to be close to each other at both frequencies [$P_\text{s}^\text{CRA} = 8.0$~mW at $f = 0.69$~THz vs $P_\text{s}^\text{CRA} = 12.5$~mW for $f = 0.297$~THz], for CRI configuration the saturation power is substantially reduced at higher $f$ [$P_\text{s}^\text{CRI} = 16.0$~mW for $f = 0.69$~THz versus $P_\text{s}^\text{CRI} = 25$~mW for $f = 0.297$~THz]. We mention that also the fitting parameters for the fits in Fig.~\ref{FigR5} satisfy the natural relation $P^{\rm CRA}_s/P^{\rm CRI}_s\simeq a^{\rm CRI}/a^{\rm CRA}$.

	\subsection{THz response of samples \#A and \#B after room light illumination}
	\label{sample_A_room}

		\begin{figure*}
		\centering \includegraphics[width=0.75\linewidth]{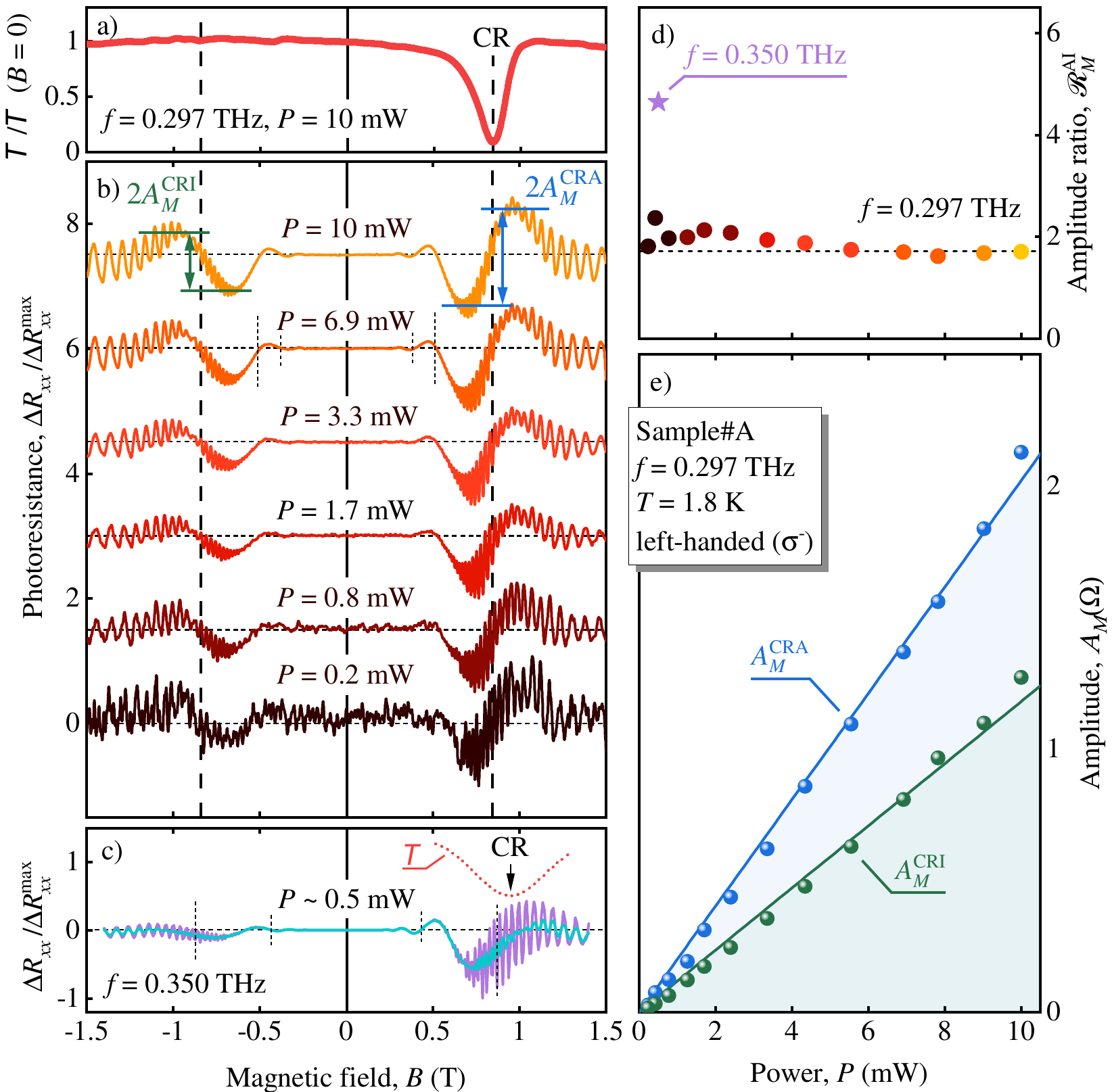}
		\caption{Results obtained at $T = 1.8$~K on sample \#A after prior exposure to room light. Panels (a), (b), and (e) show the data measured under left-handed circularly polarized radiation ($\sigma^-$) produced by the IMPATT diode operating at frequency $f = 0.297$~THz. 
  (a) Radiation transmittance recorded at radiation power of $P = 10$~mW and normalized to it's value at zero magnetic field, $\mathcal{T}/\mathcal{T}(B=0)$.
  (b) Photoresistance measured at various radiation powers. The traces are normalized to the signal's maximum, $\Delta R_{xx}/\Delta R_{xx}^{\text{max}}$ and are up-shifted for clarity. Vertical dashed lines mark the CR position for both helicities. The blue and green vertical double arrows illustrate how MIRO amplitudes, $A_M^{\text{CRA}}$ and $A_M^{\text{CRI}}$, are determined on the CRA and CRI sides. Thinner dashed vertical lines, shown as an example for the trace at $P = 6.9$~mW, indicate the second, $\pm B_{\rm CR}/2$, and third, $\pm B_{\rm CR}/3$, harmonics of the CR corresponding to the nodes of MIRO.
  (c) Photoresistance (purple trace) obtained using BWO operating at $f = 0.350$~THz and $P\simeq 0.5$~mW after exposing the sample to room light. The cyan trace depicts the photoresistance smoothed by using moving average. The red dotted curve shows the CR dip in the simultaneously measured transmittance.
  (d) Power dependence of the ratio $\mathscr{R}_M^{\rm AI}= A_M^{\text{CRA}}/A_M^{\text{CRI}}$ of MIRO amplitudes. The circles refer to the data measured at $f = 0.297$~THz [selected traces are shown in panel (b)], whereas the purple star represents the photoresistance measured at $f = 0.350$~THz [panel (c)].
  (e) Power dependence of $A_M^{\text{CRA}}$ (blue circles) and $A_M^{\text{CRI}}$ (green circles). Solid lines are linear fits. Dashed line in panel (d) shows the corresponding constant ratio $\mathscr{R}_M^{\rm AI}$.
		}
	\label{FigR2} 
	\end{figure*}

	\begin{figure*}
		\centering \includegraphics[width=0.7\linewidth]{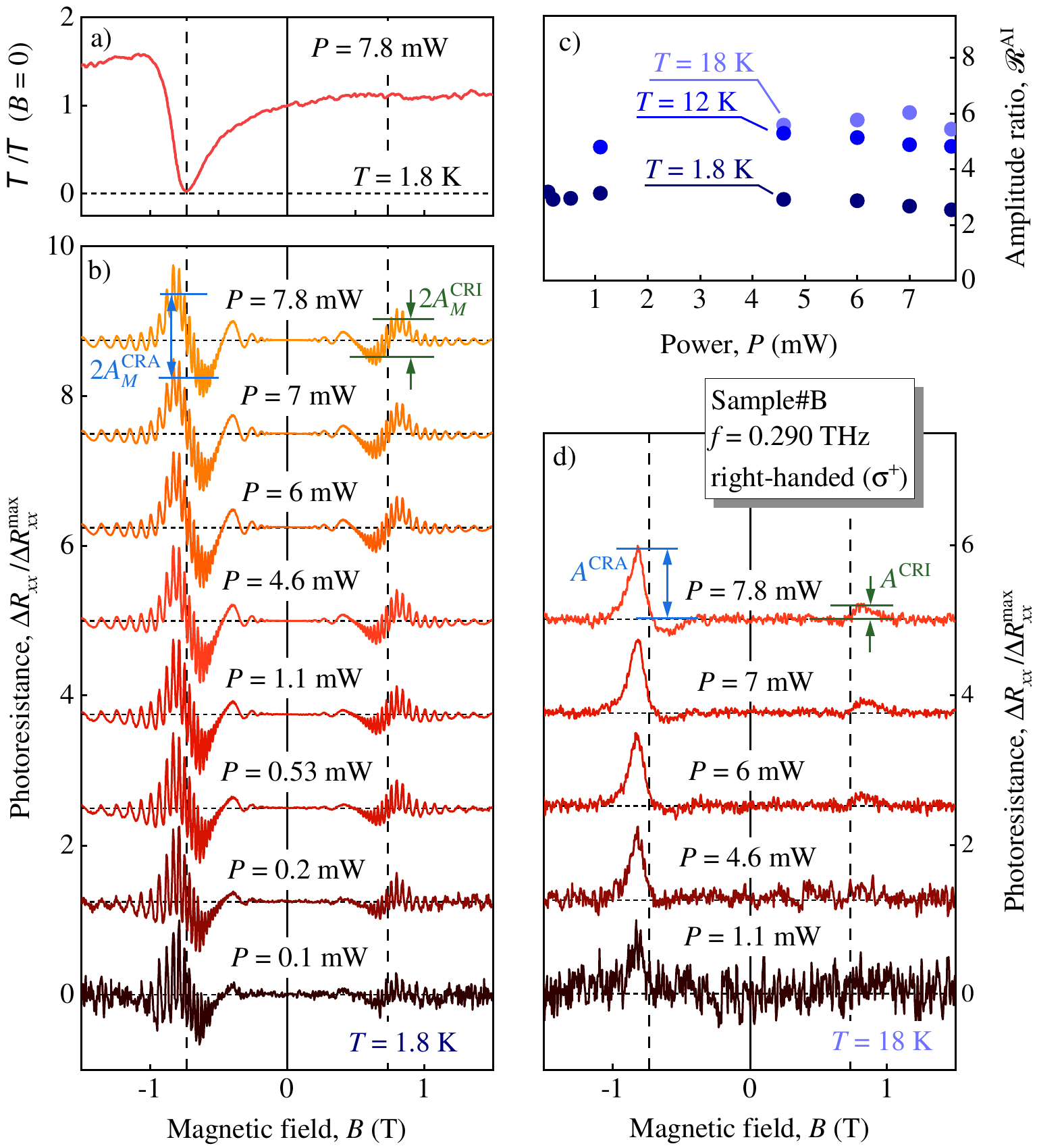}
		\caption{
  Results obtained on sample \#B after exposing it to room light and using right-handed circularly polarized radiation ($\sigma^+$) produced by the IMPATT diode operating at frequency $f = 0.290$~THz. (a) Radiation transmittance recorded at radiation power of $P = 7.8$~mW and normalized to it's value at zero magnetic field, $\mathcal{T}/\mathcal{T}(B=0)$.
    (b) and (d) Photoresistance traces measured at $T=1.8$~K and 18~K, correspondingly, and various radiation powers. The traces are normalized to the signal's maximum, $\Delta R_{xx}/\Delta R_{xx}^{\text{max}}$, and are up-shifted for clarity. Vertical dashed lines mark the CR position for both helicities. Blue and green vertical double arrows illustrate how the amplitudes are determined on the CRA and CRI sides at different $T$, see traces for $P = 7.8$~mW for $T = 1.8$ and 18~K.  
    (c) Power dependence of the ratios $\mathscr{R}_M^{\rm AI}= A_M^{\text{CRA}}/A_M^{\text{CRI}}$ of MIRO amplitudes at $T = 1.8$~K (dark blue) and power dependences of the ratios $\mathscr{R}^{\rm AI}= A^{\text{CRA}}/A^{\text{CRI}}$ of CRA and CRI peak amplitudes at $T = 12$~K (blue) and $T = 18$~K (light blue). 
    }
		\label{FigR17} 
	\end{figure*}

Figure~\ref{FigR2} shows the transmittance and photoresistance data measured with the IMPATT diode system on the same sample \#A and with the same radiation frequency $f = 0.297$~THz as in Figs.~\ref{FigR1} and \ref{FigR11}, but now obtained after brief illumination of the sample with room light prior to the measurements. Such illumination results in a substantial increase of the electron density and mobility due to the persistent photoconductivity effect~\cite{Mooney1990}, 
and also in increase of the quantum lifetime, resulting in emergence of MIRO in the photoresistance \cite{Dmitriev2012,Herrmann2016}. The behavior of the transmittance, Fig.~\ref{FigR2}(a), remains conventional after the room light illumination: It still reveals a clear CR dip at the CRA-configuration only. At the same time, the photoresistance traces, shown in Fig.~\ref{FigR2}(b) and (c), change qualitatively: Now we observe additional low-frequency $1/B$-periodic oscillations, in addition to higher-frequency SdHO-related oscillations. The former oscillations are coupled to the harmonics of the CR, yielding nodes at $B_{\rm CR}$ and its harmonics $B_{\rm CR}/2, B_{\rm CR}/3, ...$, see thick and thin dashed vertical lines in Fig.~\ref{FigR2}(b) and (c), the latter indicated for the trace at $P = 6.9$~mW. This behavior is well established for MIRO~\cite{Zudov2001,Dmitriev2012}. One observes that, similar to the bolometric photoresistance related to the electron heating, also the amplitudes of MIRO in CRA and CRI configurations are very similar, see also Fig.~\ref{FigR3} in the Appendix. For quantitative analysis, we introduce the amplitudes of MIRO, $A_M^{\rm CRA}$ and $A_M^{\rm CRI}$, as one half of the difference of $\Delta R_{xx}$ at maxima and minima around the CR position, see the upper trace in Fig.~\ref{FigR2}(b). The ratio of these amplitudes,
		\begin{equation}
  \label{ratioM}
		\mathscr{R}_M^{\rm AI} = \frac{A_M^{\rm CRA}(P)}{A_M^{\rm CRI}(P)}\,,
	\end{equation}
is plotted in Fig.~\ref{FigR2}(d). It is seen that, in contrast to the electron gas heating effects discussed in Sec.~\ref{sample_A}, the MIRO amplitudes on both sides grow linearly with $P$, Fig.~\ref{FigR2}(e). Consequently, the CRA/CRI ratio \eqref{ratioM} for MIRO is almost independent of the radiation power $P$, see Fig.~\ref{FigR2}(d). Comparing the absolute values, we see that $\mathscr{R}_M^{\rm AI}$ is about two times smaller than $\mathscr{R}^{\rm AI}$ detected at the lowest power for the bolometric photoresistance, Fig.~\ref{FigR1}(d). Figure~\ref{FigR2}(c) presents the traces obtained using the BWO system operating at $f = 0.350$~THz in sample \#A after room light illumination. Similar to the measurements at this frequency performed without prior illumination by room light, see Fig.~\ref{FigR1}(c), also for MIRO the observed ratio $\mathscr{R}_M^{\rm AI}$ is about two times larger than the highest value obtained at $f = 0.297$~THz. While the CRA/CRI ratio in the MIRO-dominated photoresistance is $P$-independent, it increases with $T$, see Fig.~\ref{FigR3} in the Appendix.
	  
Finally, we turn to the description of the results obtained in sample \#B, which features a selectively doped 16-nm GaAs QW with AlAs/GaAs superlattice barriers. Figure~\ref{FigR17} shows the transmittance and photoresistance data recorded in response to $\sigma^+$ polarized radiation in this sample (illuminated by room light prior to the measurements). As before, the transmittance curve shows a clear transmittance dip in the CRA configuration and no resonant features for the opposite magnetic field polarity. A high asymmetry of the CR dip is caused by the interference effects, see Sec.~\ref{discussion}. The photoresistance traces at different $P$  are shown for two temperatures in Figs.~\ref{FigR17}(b) and (d). At low temperature, $T=1.8$~K, the photoresistance displays MIRO modulating high-frequency SdHO-related oscillations, see Fig.~\ref{FigR17}(b). An increase of the measurement temperature results in full suppression of SdHO and in a significant suppression of MIRO. At $T=18$~K, Fig.~\ref{FigR17}(d), the photoresistance is dominated by the resonant electron heating that reduces the mobility and produces the corresponding single CR peak in the photoresistance. Figure~\ref{FigR17}(c) shows that at all $T$ the ratio $\mathscr{R}_M^{\rm AI}$ does not appreciably change with $P$, similar to the behaviour detected in sample \#A, see Fig.~\ref{FigR2}(d). Also similar to sample \#A, the ratio of the CRA and CRI signals increases at higher $T$, see also Fig.~\ref{FigR4} in the Appendix.

\section{Discussion}
\label{discussion}

Our experiments reveal that the low-temperature photoresistance of 2DES in response to circularly polarized radiation may exhibit almost symmetric CR signals at both polarities of magnetic field $B$, instead of expected single CR at a positive or negative magnetic field, depending on helicity of the incoming wave. In this work, we find that the anomalous relative amplitude of the photoresistance signals for CR active and inactive polarities of $B$, the ratio $\mathscr{R}^{\rm AI} = A^{\rm CRA}/A^{\rm CRI}$ introduced in Eq.~\eqref{ratio}, is highly sensitive to both radiation power and temperature, see Figs.~\ref{FigR1}, \ref{FigR11}, \ref{FigR5}, and \ref{FigR17}, also supported by additional data in the Appendix. While the CRI response in the photoresistance is anomalous, the simultaneously measured transmittance displays an ordinary behavior with a single CR dip. Furthermore, the photoresistance on the CRA side can be pretty well captured by rather common and well-established mechanisms related to the electron gas heating \cite{Ganichev2005} and MIRO \cite{Dmitriev2012}.

Therefore, in the analysis below, we first review such conventional behavior of the transmittance and photoresistance under a uniform circularly polarized wave in terms of the local dynamic and static conductivity of a uniform isotropic 2DES. This will include electrodynamic effects, such as CR reflection, absorption, and transmittance of a high-mobility 2DES in Faraday geometry in the presence of the Fabry-P\'{e}rot interference in the substrate, and electron heating effects in the photoresistance. The latter should be distinguished in the classical regime of $\mu$-photoconductivity (hot electron bolometer) and in the quantum regimes of SdHO and MIRO, as well as in the linear and nonlinear regimes with respect to radiation power. Application of this theory to the regular transmittance and to the photoresistance on the CRA side in our experiments allows us to estimate the expected ratio $\mathscr{R}^{\rm AI}$ (thus providing a quantitative description of the helicity anomaly) and to analyze the observed power dependence of $\mathscr{R}^{\rm AI}$. After that we discuss possible microscopic origins of the helicity anomaly.

\subsection{CR transmittance, absorptance, and  reflectance of a high-mobility 2DES}

\begin{figure*}
	\centering \includegraphics[width=0.6\linewidth]{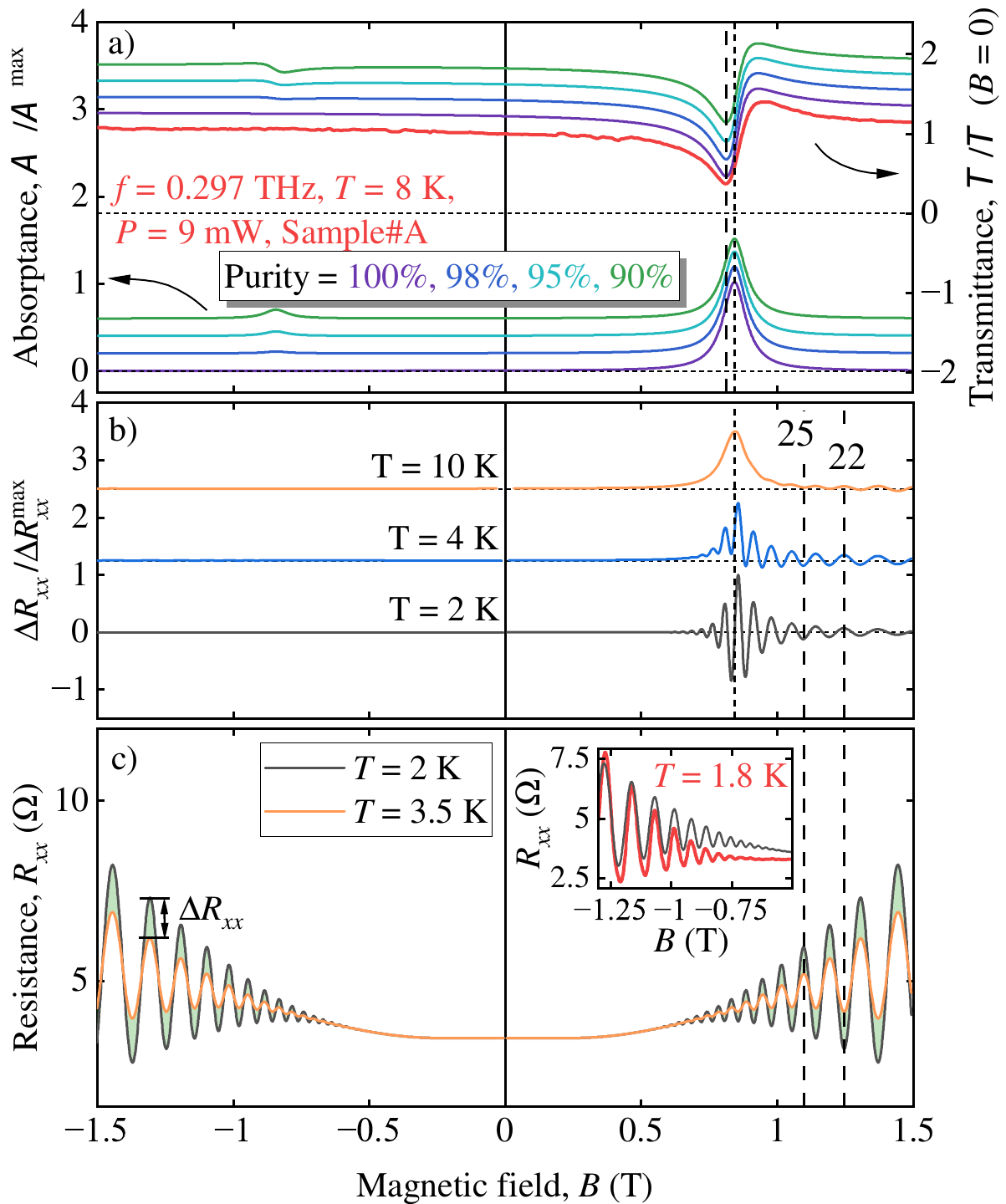}
	\caption{(a)  Drude transmittance $\mathcal{T}^\text{D}(B)$ and absorptance ${\cal A}^\text{D}(B)$ calculated using Eqs.~\eqref{TD} and \eqref{AD}, with parameters $B_\text{CR}=0.82$~T and $\beta=0.023-i0.023$~T determined from the transmittance measured at $f = 0.297$~THz, $T = 8$~K, $P = 9$~mW on sample \#A (red). Theoretical curves, calculated for the circular polarization purity of 100\%, 98\%, 95\%, and 90\% (bottom to top) are shifted by a constant offset for better visibility. Minimum of transmittance at $B_\text{CR}$ and maximum of absorptance at $B_\text{CR}- \text{Im} \beta$ are marked by vertical dashed lines. (b) Normalized photoresistance, $\Delta R_{xx}/\Delta R_{xx}^{\text{max}}$, calculated according to Eq.~\eqref{dRlinear} and \eqref{eq:lk} for $T=2$, $4$, and $10$~K.  (c) Dark magnetoresistance, $R_{xx}(B)$, calculated using Eq.~\eqref{eq:lk} for $T=2$ and $3.5$~K. Vertical dashed lines in (b) and (c) at filling factors $\nu=$22 and 25 illustrate the relative phase of SdHO in hotoresistance and dark magnetoresistance. Inset in (c) shows measured (red) and calculated (black) magnetoresistance for sample \#A and $T = 1.8$~K. The fit yields parameters $R_0 = 3.4$~$\Omega$, $\tau_\text{q}=1.3$~ps, and $B_\text{inh}=1.1$~T used for calculations in (b) and (c).
  }
	\label{FigR13_2} 
\end{figure*}

For a normal incidence of a circularly polarized monochromatic wave to the surface of a sample, containing a 2DES and placed in a perpendicular magnetic field $B=B_z$ (Faraday geometry), the transmittance $\mathcal{T}(B)$ is given by the standard expression, $\mathcal{T}(B)=4/|s_1(1+Z_0\sigma_\eta )+s_2|^2$ \cite{Abstreiter1976,Chiu1976,Mikhailov2004,Dmitriev2012,Savchenko2021}. Here $\sigma_\eta=\sigma_{xx}+i\eta\sigma_{yx}$ is the complex dynamic conductivity of a uniform isotropic 2DES, $\eta=\pm 1$ denotes the helicity of the wave, $Z_0\approx 377$~$\Omega$ is the impedance of th efree space, and the complex parameters $s_{1,2}=\cos\phi-i n_r^{\mp 1}\sin\phi$ describe the Fabry-P\'{e}rot interference due to multiple reflections in the GaAs substrate. The interference phase $2\phi=4\pi w/\lambda_r$ is given by the ratio of the sample thickness $w$ and the wavelength $\lambda_r=c/f n_r$, where $n_r\simeq 3.6$ is the refractive index of the GaAs substrate. With $\sigma_\eta$ taken in the classical Drude form, $\sigma^\text{D}_\eta=e^2 n_e/(\mu^{-1}-i B_\text{CR}+i\eta B)$, where $B_\text{CR}=2\pi f m_\text{CR}/e$ denotes the position of the CR, the resulting Drude transmittance $\mathcal{T}^\text{D}$ and corresponding Drude absorptance, ${\cal A}^\text{D}=Z_0 \mathcal{T}^\text{D} \mathrm{Re}\,\sigma^\text{D}_\eta$, are given by
\begin{equation}
	\label{TD}
	\mathcal{T}^\text{D}(B)=|\alpha|^2\left|1-\dfrac{\beta}{\mu^{-1}+\beta - i B_\text{CR}+i\eta B}\right|^2,
\end{equation}
\begin{equation}
	\label{AD}
	{\cal A}^\text{D}(B)=\dfrac{Z_0|\alpha|^2 e^2 n_e/ \mu}{|\mu^{-1}+\beta - i B_\text{CR}+i\eta B|^2}. 
\end{equation}
Here $\alpha=2/(s_1+s_2)$ and $\beta=e n_e Z_0/(1+s_2/s_1)$. In the case of illumination of the sample from the 2DES side, as in Figs.~\ref{FigR1}(c) and \ref{FigR2}(c), the expression \eqref{AD} for ${\cal A}^\text{D}$ should be additionally multiplied by the factor $|s_1|^2$. 

Typical shapes of normalized $\mathcal{T}^\text{D}(B)$ and corresponding ${\cal A}^\text{D}(B)$, calculated according to Eqs.~\eqref{TD} and \eqref{AD}, are shown in Fig.~\ref{FigR13_2}(a) together with the experimental transmittance trace (red line) recorded at $f = 0.297$~THz, $T = 8$~K, and $P = 9$~mW on sample \#A, without prior exposure to room light. The shape of measured $\mathcal{T}(B)$ is well reproduced using $\eta=+1$, corresponding to the left-handed circularly polarized radiation ($\sigma^-$), the transport CR width $\mu^{-1}=0.03$~T obtained using $\mu$ extracted from magnetotransport measurements (see Table \ref{transport_para}), the CR position $B_\text{CR}=0.82$~T corresponding to the cyclotron mass $m_\text{CR}=0.07 m_0$, and with $\beta=0.023-i0.023$~T. The latter parameter accounts for strong metallic reflection from 2DES (superradiant decay \cite{Zhang2014}), modified by the Fabry-P\'{e}rot interference \cite{Savchenko2021}, and will be addressed in more detail below. 

Most importantly, the conventional Drude theory combined with Maxwell equations accurately reproduces the observed shape of the transmittance for 100~$\%$ $\sigma^-$ radiation, confirming the purity of circular polarization in the transmitted, and, thus, also in the incoming wave. For comparison, in Fig.~\ref{FigR13_2}(a) we also plot $\mathcal{T}^\text{D}(B)$ for a plane wave with small admixtures (2, 5, and 10 $\%$) of $\sigma^+$ circular component. It is seen that even a few percent admixture is clearly visible in the transmittance on the CR-passive side $B<0$. On the other hand, the measured transmittance data do not show any traces of such admixture above the noise level, see also Figs.~\ref{FigR1}, \ref{FigR11}-\ref{FigR17}, and Appendix. 
This observations provide a clear evidence that the anomalous CRI signals observed in the photoresistance cannot originate from an admixture of the opposite helicity components in the incoming wave. 

A more detailed inspection of the transmittance traces shows that their shape and width vary significantly with the radiation frequency. These modifications mainly come from variations of the Fabry-P\'{e}rot interference parameters $s_{1,2}$ which rapidly change with the radiation frequency. In particular, both for constructive ($s_{1,2}=\cos\phi=1$) and destructive ($\sin\phi=1$, $s_{1,2}=-i n_r^{\mp 1}\simeq -i 3.6^{\mp 1}$) Fabry-P\'{e}rot interference the CR dip in the transmittance has a symmetric form, but in the latter case the parameter $\Re[\beta]=\Re[e n_e Z_0/(1+s_2/s_1)]$ defining the radiative width of the CR is about 14 times smaller. For sample \#A without prior exposure to room light (the electron density $n_e=7\times10^{11}$~cm$^{-2}$, see Table \ref{transport_para}), one obtains $\beta=0.21$ for constructive and $\beta=0.015$ for destructive interference. This substantial difference of the CR width can be seen, for instance, in the transmittance traces presented in Fig.~\ref{FigR1}(a) (destructive interference) and Fig.~\ref{FigR1}(c) (constructive interference), obtained on the same sample at different frequencies. Furthermore, as illustrated in Fig.~\ref{FigR13_2}(a), intermediate values of the interference phase (corresponding to $\rm{Im}\beta\neq 0$) give rise to an asymmetric shape of the transmittance and a corresponding shift of the maximum of the absorptance (dashed lines). Such asymmetric shape of the measured $\mathcal{T}(B)$ is also seen, e.g., in Figs.~\ref{FigR5}(a) and \ref{FigR17}(a). As follows from Eqs.~\eqref{TD} and \eqref{AD}, these changes are controlled by the imaginary part of the interference parameter $\beta$. In contrast  to $\mathcal{T}(B)$, the shape of the Drude absorption remains Lorentzian for all values of $\beta$.

Using Eq.~\eqref{AD}, we can estimate the expected Drude CRA/CRI ratio of the absorbed power, assuming the 100\% circularly-polarized radiation that remains uniform in the plane of the 2DES. Taking for definiteness the case of helicity $\eta=1$, to obtain this ratio we should relate the value of absorptance at the maximum of the Lorentzian, $B=B_\text{CR}-{\rm Im}\beta$ (CRA), to its value at the opposite magnetic field (CRI). Since the CR width is maximal for the condition of constructive interference, the minimal CRA/CRI ratio for the Drude absorption can be estimated as
\begin{equation}
    \label{Rmax}
    \mathscr{R}_\text{min}^{\rm AI}=\dfrac{(\mu^{-1}+\beta)^2+4 B_\text{CR}^2}{(\mu^{-1}+\beta)^2}\simeq 47,
\end{equation}
where we used the values $\mu^{-1}=0.03$~T, $B_\text{CR}=0.82$~T, and $\beta=0.21$~T quoted above. At different conditions the ratio can be much larger, for example, for destructive interference, $\beta=0.015$~T, one obtains $\mathscr{R}_\text{max}^{\rm AI}\simeq 1330$. 

Importantly, these values were estimated for 100\% circular polarization, in which case one does not expect any resonant absorption in the CRI regime, see Fig.~\ref{FigR1}(a), in contrast to our findings for the photoresistance, where strong {\em resonant} CRI features are observed. This suggests that in reality the electromagnetic field acting on the 2DES is not a circularly polarized uniform plane wave. On the other hand, as we have seen, the transmittance data provide a clear evidence that the purity of circular polarization in both incoming and transmitted wave is rather high. The estimate \eqref{Rmax} and Fig.~\ref{FigR1}(a) show that in our samples a several percent admixture of the wave with opposite helicity should lead to a similar CRI signal in the photoresistance, also not exceeding several percent of the CRA signal.
Therefore, our observations of the anomalously low CRA/CRI ratio in the photoresistance
suggest the importance of near-field effects, which are discussed below in Sec.~\ref{SecNearField}. Before that, we shortly review the well established links between the CR absorptance and photoresistance in a uniform EM field which, in particular, explain the observed nonlinear effects in the photoresistance.

\subsection{Electron transport and photoresistance due to electron heating effects}

At the CRA side, photoresistance traces (see Figs.~\ref{FigR1}, \ref{FigR11}, and \ref{FigR5}) display the well established behavior associated with resonant electron gas heating under the CR absorption. Under continuous illumination, the heating effects are quite generally described using the energy balance equation,
\begin{equation}
	\label{EBE}
	{\cal A}(I) I= Q(I),
\end{equation}
which expresses the stationary condition that the radiation energy ${\cal A}(I) I$ absorbed by electrons is fully compensated by the energy flow $Q(I)$ from hot electrons to the lattice \cite{Gantmakher1988,Ganichev2005}. The lattice serves as a thermal bath and can be usually assumed to remain at the measurement temperature $T$. As discussed below, in general both the absorptance ${\cal A}$ and losses $Q$ depend on the intensity $I$ of incoming radiation. Importantly, at low electron temperatures, corresponding to conditions of our experiments, the equilibration of the absorbed radiation energy within the electron system, governed by the inelastic electron-electron scattering, is much faster than the rate of energy transfer to the lattice. Therefore, the stationary non-equilibrium energy distribution of electrons under continuous illumination still has the shape of equilibrium Fermi-Dirac distribution, but with the measurement temperature $T$ replaced by an elevated electron temperature $T_\text{e}>T$. On the other hand, the heating-induced changes of the chemical potential in a degenerate high-density 2DES can be safely ignored. Thus the electron temperature $T_\text{e}$ is well defined and fully characterizes the electron gas heating effects. The value of $T_\text{e}$ should be found self-consistently from the energy balance equation \eqref{EBE}. 

In the simplest case of linear heating, corresponding to the limit of low $I\to 0$, one replaces ${\cal A}(I)$ with the $I$-independent linear-response absorptance, ${\cal A}(I\to 0)$, given in our case by Eq.\eqref{AD}. In turn, the energy losses $Q$ can be rewritten as $(T_\text{e}-T)\tau_\text{e-ph}^{-1}$, where $\tau_\text{e-ph}^{-1}(T_\text{e}, T)$ characterizes the electron-phonon energy relaxation rate. Therefore, in the linear heating regime $T_\text{e}-T\propto I {\cal A}^\text{D}(B) \tau_\text{e-ph}(T, T)$, where only the absorption rate essentially depends on $B$. Note that in the linear regime $(T_\text{e}-T)/T\propto I\to 0$ the inelastic relaxation rate, $\tau_\text{e-ph}^{-1}(T_\text{e}, T)$, that is usually strongly temperature-dependent, and, in general, depends in a distinct way on electron and lattice temperatures, should be replaced by the intensity independent rate $\tau_\text{e-ph}^{-1}(T, T)$. 
In turn, the photoresistance due to electron heating in the linear regime is given by
\begin{equation}\label{dRlinear}
	\Delta R= \dfrac{\partial R}{\partial T_\text{e}}(T_\text{e}-T).
\end{equation} 
The dependence of the resistance $R(T_\text{e},T)$ on the electron and lattice temperature is also in general different, so it can be not sufficient to measure the $T$-dependence of dark resistance to determine the coefficient $\partial R/\partial T_\text{e}$. 

As outlined above in Sec.~\ref{sample_A}, in our experiments the electron heating produces a combination of two distinct effects in the photoresistance. Namely, at high $T$ we observe a single CR peak in $\Delta R$ caused by heating-induced decrease of the electron mobility \cite{Ganichev2005, Muraviev2013, Freitag2012, Betz2012, Heyman2015, Ryzhii2019, Jago2019},
while at the lowest $T$ the main effect of the resonant electron heating comes from the reduction of the SdHO amplitude reflecting their exponential sensitivity to the electron temperature $T_\text{e}$. 

In the high-$T$ regime where the SdHO are absent, the longitudinal resistance is $B$-independent and is given by the Drude expression,  $R_0=g/(e n_e \mu)$, where $g$ is a geometrical factor depending on the type of measurements. According to Eq.~\eqref{dRlinear}, in this case the $B$-dependence of the linear photoresistance $\Delta R\propto-I({\partial \mu}/{\partial T_\text{e}}) {\cal A}^\text{D}(B)$ should directly reproduce the Lorentzian shape of the magnetoabsorptance, see Eq.~\eqref{AD} illustrated in Fig.~\ref{FigR13_2}(a). Our observations in the CRA regime at high $T$ are well captured by this mechanism termed $\mu$-photoconductivity, see  high-$T$ traces in Fig.~\ref{FigR11}. The positive sign of $\Delta R$ is consistent with the decrease of mobility at high $T_\text{e}$ due to acoustic phonon scattering. We emphasize that so far we are discussing the linear regime of weak heating, $\Delta R\propto I$. The nonlinear effects will be considered in next subsection.

At lower $T$, one should account for the combined effect of CR heating on the  mobility and SdHO, as illustrated in panels (b) and (c) of Fig.~\ref{FigR13_2}. In panel (c), the calculated resistance in the absence of radiation is shown for two temperatures, $T=2$~K and $T=3.5$~K. Here we use standard Lifshitz-Kosevich formula for SdHO~\cite{Shoenberg1984,Ando1982,Dmitriev2012},
\begin{equation}
	\label{eq:lk}
	R_{xx}(B) =R_0 +2 R_0\delta^2- 4 R_0 \delta_\text{inh}\delta \frac{\,\text{X}(T)}{\sinh \text{X}(T)}\cos(\pi \nu).
\end{equation}
The SdHO, described by the last term, are $1/B$-periodic oscillations caused by the modulation of the density of states, and their period is controlled by the carrier density $n_e$ (the filling factor $\nu= 2\pi\hbar n_e/e |B|$). At zero temperature, $T=0$, the decay of SdHO at low $B$ is described by the Dingle factor $\delta  = \exp(-\pi/\omega_\text{c} \tau_\text{q})$, where the quantum relaxation time $\tau_\text{q}$ characterizes the disorder broadening of Landau levels separated by $\hbar\omega_\text{c}=\hbar e |B|/m$. The factor containing $\text{X}(T) = 2\pi^2 k_\text{B} T_\text{e}/\hbar \omega_\text{c}$ accounts for the additional $T$-smearing. For generality, in Eq.~\eqref{eq:lk} we also include the non-oscillating quantum correction $2 R_0\delta^2$ and an additional damping factor, $\delta_\text{inh}=\exp(-B_\text{inh}^2/B^2)$, accounting for possible smooth fluctuations of the filling factor across the sample~\cite{Dmitriev2012}.  In panel (c) of Fig.~\ref{FigR13_2}, the parameters entering Eq.~\ref{eq:lk} (see caption) are chosen such that the calculated $R_{xx}$ closely reproduces the experimental magnetoresistance, as illustrated in the inset.

Using Eq.~\ref{eq:lk} with the chosen parameters together with Eq.~\ref{dRlinear}, in panel (b) of Fig.~\ref{FigR13_2} we illustrate the corresponding changes of the photoresistance behavior with the measurement temperature at low intensities. It is seen that the photoresistance oscillations exhibit maxima (minima) at even (odd) Landau level filling factors, unlike SdHO in the dark magnetotransport having minima (maxima) at even (odd) filling factors, see dashed lines in panels (b) and (c). This behavior is well reproduced in our photoresistance data for all samples in conditions where the SdHO-related heating mechanism dominates, and corresponds to the expected reduction of the oscillation amplitude due to heating. At $T=2$~K, the SdHO-related contribution to the photoresistance in Fig.~\ref{FigR13_2}(b) dominates and the oscillations are symmetric with respect to $\Delta R=0$; the envelope is distorted from the Lorentzian shape due to the exponential decay of SdHO at low $B$. At higher temperatures, $T=4$ and $10$~K in Fig.~\ref{FigR13_2}(b), the SdHO get gradually suppressed, and the non-oscillating contribution related to $\mu$-photoconductivity takes over. This leads to the Lorentzian shape of the photoresistance reproducing that of the magnetoabsorption. All these general features are clearly seen in our photoresistance traces, see Figs.~\ref{FigR1}(b) and (c), \ref{FigR11}(b), and \ref{FigR5}(b).

\subsection{Nonlinearity in CRA}

Our experiments reveal that the increase of the radiation intensity leads to a sublinear power dependence of the photoresistance, see, e.g., Figs.~\ref{FigR1}(e) and \ref{FigR5}(d). We attribute these changes to the nonlinear electron heating. In general, the nonlinearity of $T_\text{e}(I)$ can be caused by the saturation of the radiation absorption ${\cal A}(I)$ and/or the nonlinearity of energy losses $Q(I)$, see Eq.~\eqref{EBE}. The former usually becomes essential at very high intensities $I\gtrsim 100$ kW/cm$^2$ \cite{Helm1985,Rodriguez1986,Mics2015,Candussio2021a} and thus plays no role here. By contrast, the nonlinear dependence of the rate of energy losses $Q(I)$ becomes essential at much lower powers, as soon as the heating $T_\text{e}-T$ is no longer much smaller than $T$. One of the best known examples of such nonlinearity is the case of low-angle electron-phonon scattering which was considered theoretically in Ref.~\cite{Gantmakher1988} and observed in saturation of THz photoresistance in Si-MOSFET structures \cite{Ganichev2005,Beregulin1988,Beregulin1990}. At the lowest $T$, the dependence  $Q\propto T_\text{e}^5-T^5$ is very strong, resulting in fast growth of energy losses with increasing $I$, and, consequently, to sublinear growth (saturation) of the electron temperature $T_\text{e}$. At higher $T$, the dependence usually becomes weaker, and saturation  takes place at higher intensities. Finally, at very high $T$, the universal regime of $Q\propto T_\text{e}-T$, characterized by equal thermal occupation of all involved phonon modes, is reached. In this case, the nonlinearity of the energy losses is absent even at high $I$, and possible nonlinearities of the photoresistance are governed by other mechanisms \footnote{Apart from saturation of the radiation absorption ${\cal A}(I)$ mentioned above, these can result from details of the temperature dependence of resistance $R(T_\text{e},T)$, which can be governed by different microscopic mechanisms in different temperature regimes. In other words, at high $I$ Eq.~\eqref{dRlinear} is not always applicable, and should be replaced by a more general expression for the photoresistance, $\Delta R= R[T_\text{e}(I),T]-R[T_\text{e}(I\to 0),T]$.} In addition to the above considerations, the basic requirement for the nonlinearity of the electron heating is that the resulting $(T_\text{e}-T)/T$ is not too small. The higher is the measurement temperature, the higher radiation intensity is required to fulfill this  condition. This agrees well with our observations at high $T$ showing no saturation and, correspondingly, no essential changes of the CRA/CRI ratio with the radiation intensity.

Apart from the sublinear growth of the magnitude of resonant photoresistance signals with intensity, clearly seen in Figs.~\ref{FigR1}(e) and \ref{FigR5}(d), the nonlinear electron heating naturally leads to an additional broadening of these signals. Indeed, collecting the photoresistance signals obtained at different intensities and normalizing them to the radiation power, in Fig.~\ref{FigR13} we reveal the strongest saturation near the maxima of the CRA signals, corresponding to the maximal absorption and electron heating. By contrast, both at the wings of the CRA signals and on the CRI side the electron heating remains almost linear even at the highest $I$. This explains the observed increase of the CRA/CRI ratio \eqref{ratio} at low intensities as a result of a transition from the nonlinear to the linear electron heating regime. However, even at the lowest $I$ available for our measurements this ratio remains anomalously low, see Figs.~\ref{FigR1}(d) and \ref{FigR5}(c). The situation becomes different at higher $T$, as illustrated by a similar analysis of the shape of the photoresistance at $T=6$~K, see Fig.~\ref{FigR13bis} in the Appendix. As discussed above, at higher $T$ the nonlinear heating effects become weaker. Accordingly, the shape of the CR photoresistance and, therefore, also the CRA/CRI ratio at high $T$ do not appreciably change within the available range of radiation power, see also  Fig.~\ref{FigR13_3} in Appendix.

\subsection{CRI photoresistance and nature of the anomalous CRI absorption }
\label{SecNearField}

The discussion above demonstrates that both transmittance and photoresistance data obtained in the CRA regime are conventional and are well described by the standard theory. At the same time, the strength of the detected CRI signals in the photoresistance is clearly anomalous. 
Taking into account the very similar shape and comparable magnitude of CRA and CRI signals, see Figs.~\ref{FigR1}(b) and (c), \ref{FigR11}(b), and \ref{FigR5}(b), the whole photoresistance data could be readily explained assuming them as a response to elliptical, and sometimes, almost linearly polarized radiation. This, however, contradicts the measured transmittance, Figs.~\ref{FigR1}(a) and (c), \ref{FigR11}(a), and \ref{FigR5}(a), which unambiguously demonstrates a high purity of the incoming circularly polarized radiation.

The central assumption underlying this apparent paradox is that conventional theory of the CR in transmittance, absorptance, and photoresistance considers a uniform local current response of 2DES to the electric field of a uniform plane circularly polarized wave~\cite{Dmitriev2012,Savchenko2022}. Such response is fully encoded in the local dynamic conductivity, and becomes resonant at positive or negative $B$ only, depending on the helicity of the incoming wave. It is clear that this standard and widely employed theory is incompatible with the observed anomalously strong CRA absorption.

That is why in Ref.~\cite{Moench2022b} it was proposed that explanation of the helicity anomalies should necessarily involve a mechanism of conversion of the uniform THz radiation into non-uniform evanescent near-fields, which are present only in close vicinity of the 2DES and are accompanied by the emergence of spatially non-uniform electric currents there. Unlike the uniform field of the incoming plane wave, the near-field couples to longitudinal plasmonic excitations. Therefore, the polarization state is inevitably altered and is different from that of the incoming wave. In Ref.~\cite{Moench2022b} it was conjectured that the electron response to the THz electric field can be essentially modified near rare strong scattering centers or inhomogeneities~\cite{Dmitriev2012,Baskin1978,Bobylev1995,Mirlin2001,Dmitriev2004,Beltukov2016,Dorozhkin2017,Chepelianskii2018} mediating a near-field coupling between the two helicity modes and thus enabling the polarization immunity. Alternatively, near-field effects may originate from scattering of the plane EM wave itself, i.e., on charged or dipole centers in the dielectric matrix surrounding the 2DES. Irrespective the microscopic origin of the near-fields, their scattered components of the wave can indeed produce similar CRA and CRI signals in the photoresistance, superimposed on the regular CRA response from the plane wave component. Within this scenario, the controversy between the helicity dependence of measured transmittance and photoresistance is immediately resolved. Indeed, the magnetotransmittance is measured in the far field, and thus its shape cannot be affected by near-field effects as long as the far-to-near-field conversion depends weakly on the applied magnetic field. In contrast, the absorption by the 2DES and, consequently, the photoresistivity are directly sensitive to both far- and near-field components. 

It is still remarkable that the detected CRA and CRI signals in the photoresistance turn out to be comparable in magnitude despite high quality and uniformity of the studied 2DES. Indeed, usually the near-field effects are considered to arise due to a presence of macroscopic metallic objects~\cite{Zayats2009,Keller2012,Girard1996}
while in present experiments we deal with a nominally uniform 2DES with the size much larger than the laser spot where strong near-field effects are usually not expected to occur. However, it should be taken into account that a significant part of the incoming radiation on the CRA side is reflected from the 2DES. This metallic reflection strongly suppresses the plane wave component reaching the 2DES at the CR and, therefore, also the scattered near-fields emerging on the CRA side. By contrast, on the CRI side the circularly polarized plane wave is almost fully transmitted. Thus, the generated near-fields in the CRI regime can be much stronger than on the CRA side. 

Furthermore, in the present work we demonstrate that the CRA/CRI ratio $\mathscr{R}^{\rm AI}$, Eq.~\eqref{ratio}, is sensitive to the radiation power, see Figs.~\ref{FigR1}(d) and \ref{FigR5}(c). It grows with lowering power and reaches values considerably larger than unity at low powers. Thus, the most puzzling {\em complete polarization immunity} is observed only at high powers, where it can be readily explained as a result of nonlinear electron heating, as outlined above. These observations are also in agreement with previous reports of complete polarization immunity in ultra-high power measurements of the photoresistance response to pulsed THz radiation~\cite{Herrmann2017}.  As discussed in Sec.~\ref{sample_A}, the intensity dependence of the CRA/CRI ratio originates from different absorption strengths in the CRA and CRI regimes, which translate to different saturation powers $P_s$ entering Eq.~\eqref{saturation}. For instance, analysis of the data shown in Fig.~\ref{FigR1}(d) gives a factor of two for the ratio $P_s^{\rm CRI}/P_s^{\rm CRA}$ of saturation powers, consistent with the ratio of the linear coefficients, $a^{\rm CRA}/a^{\rm CRI}$. Consequently, at high power the extrapolated ratio $\mathscr{R}^{\rm AI}$ in Eq.~\eqref{saturation} approaches unity. Comparing the photoresistance data obtained at two frequencies, see Figs.~\ref{FigR1} and \ref{FigR5}, we have found that the saturation powers $P_s$ are similar in both cases, while the corresponding saturation intensity is several times larger at higher frequency. Such increase is expected from the $f^{-2}$ scaling of the Drude absorptance.
Taken together, these observations suggest that, independent of the nature of the absorption (near-field vs. plane wave), the nonlinearity of the electron gas heating is observed when the absorbed radiation energy per unit area of the 2DES exceeds a certain value. This provides an additional evidence that the observed nonlinearities in the photoresistance are primarily associated with the nonlinearity of the energy losses $Q(I)$ which should be insensitive to the magnetic field as well as polarization and nature of the absorbed radiation.

Finally, we briefly comment on the observed anomalous helicity dependence of MIRO, see Figs.~\ref{FigR2} and \ref{FigR17}. In contrast to the resonant photoresistance caused by electron gas heating, MIRO amplitudes showed no saturation at elevated radiation powers. Even at the lowest temperature ($T=1.8$~K) they scale linearly both in CRA and CRI regimes, see Fig.~\ref{FigR2}(e). Correspondingly, the found CRA/CRI ratios $\mathscr{R}^{\rm AI}_M$ for MIRO, varying in the range from approximately two to six depending on the sample and the experimental condition, were also almost independent of $P$, see Figs.~\ref{FigR2}(d) and \ref{FigR17}(c). These findings are not surprising and are in line with the results of previous studies of this effect, since the mechanisms of MIRO are not related to electron gas heating. These oscillations result from resonant absorption between neighboring or distant Landau levels broadened by disorder, which leads to an oscillatory correction to the energy distribution of electrons (inelastic mechanism) and spatial displacements of electron orbits in the direction determined by the ratio of radiation and cyclotron frequencies (displacement mechanism)\cite{Dmitriev2012}. At the same time, similar to the resonant photoresistance in the linear regime with respect to radiation power, MIRO are proportional to the radiation absorption. Therefore, the CRI absorption induced by near-field effects described above should affect the polarization dependence of MIRO in the same way as for the photoresistance induced by electron gas heating.
Our studies indeed reveal similar CRA/CRI ratios for both the bolometric effect and MIRO, confirming that the anomalous helicity dependence of MIRO just reflects the anomalous resonant CRI absorption, and is not an inherent property of this effect~\cite{Savchenko2022}.

	\section{Summary}
Summarizing, in our work we demonstrate that the recently observed helicity anomaly in resonant photoresistance, induced by the CR heating of the electron gas, possesses strong dependence on the power of the circularly polarized radiation. Namely, we observe that the CRA/CRI ratio in the photoresistance grows with lowering power, and show that this behavior is associated with the saturation of the resonant electron gas heating, characterized by different saturation powers for CRA and CRI magnetic field polarities. Importantly, even at lowest powers the  CRA/CRI ratio remains anomalously low. The analysis of our results reveals that the overall behavior of the photoresistance is well captured by conventional theory including electrodynamic effects, such as strong metallic reflection from the 2DES in the region of the CR and the Fabry-P\'{e}rot interference in the substrate, combined with the standard theory of the electron gas heating. However, to get such an agreement, one needs to assume some source of resonant absorption in the CRI regime. This suggests the presence of scattered near-fields in the vicinity of the 2DES, which by their nature have drastically different polarization properties as compared to the incoming circularly polarized plane wave, and thus can indeed produce a resonant CR response for both polarities of the magnetic field. While the observed anomalous behavior of the CR absorption and corresponding nonlinear photoresistance are of interest in their own right, the presented results are of importance for any polarization-sensitive photoelectric studies in 2DES.

	\section{Acknowledgements}
	
    We thank Imke Gronwald for the support in the sample fabrication. We acknowledge the financial support of the Deutsche Forschungsgemeinschaft (DFG, German Research Foundation) via Project-ID 314695032 – SFB 1277 (Subprojects A01, and A04) and via grant DM~1/6-1 (I.A.D.), and of the Volkswagen Stiftung Program (97738).

	\appendix
	\section*{Appendix}
	
\begin{figure*}
		\centering \includegraphics[width=0.8\linewidth]{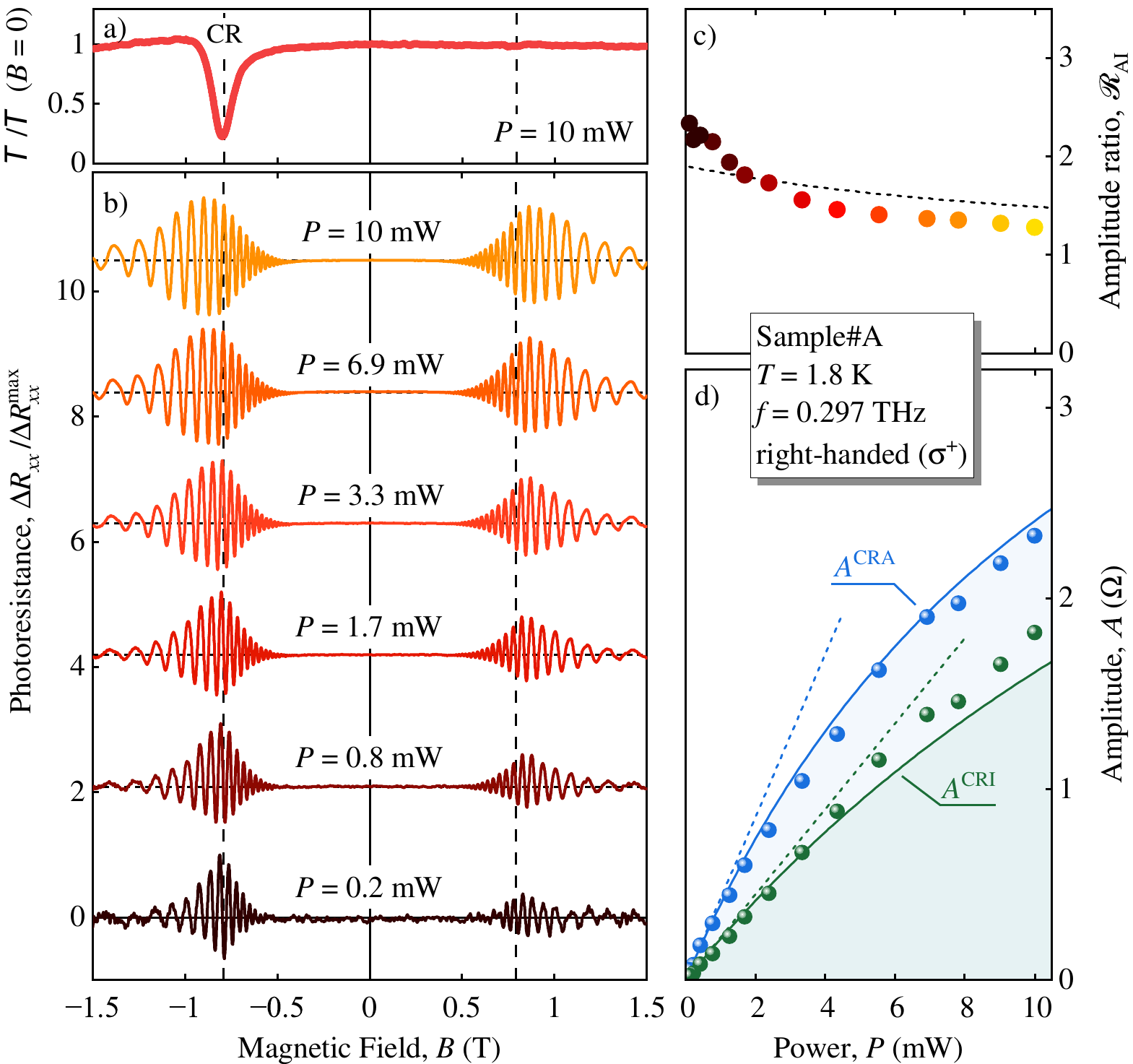}
		\caption{Results obtained in sample \#A at $T=1.8$~K without prior room light illumination and subjected to
  right-handed circularly polarized radiation ($\sigma^+$) produced by the IMPATT diode operating at frequency $f = 0.297$~THz. 
  (a) Radiation transmittance recorded at radiation power of $P = 10$~mW and normalized to its value at zero magnetic field, $\mathcal{T}/\mathcal{T}(B=0)$. 
  (b) Photoresistance measured at various radiation powers. The traces are normalized to the signal's maximum, $\Delta R_{xx}/\Delta R_{xx}^{\text{max}}$ and are up-shifted for clarity. The position of the CR is indicated by vertical dashed lines for both helicities. (c) Power dependence of the ratio $\mathscr{R}^{\rm AI}= A^{\text{CRA}}/A^{\text{CRI}}$ of the CRA and CRI amplitudes. (d) Power dependencies of $A^{\text{CRA}}$ (blue) and $A^{\text{CRI}}$ (green), as extracted from unnormalized photoresistance data $\Delta R_{xx}$ measured at $f = 0.297$~THz. Solid lines are calculated according to Eq.~\eqref{saturation} using $a$ and $P_s$ as fitting parameters. Dashed lines illustrate the linear part $A=a P$ of the corresponding fits (
  $a^\text{CRA} = 0.43$~$\Omega$/mW and $P_s^\text{CRI} = 12.9$~mW at the CRA side,  
  $a^\text{CRI} = 0.224$~$\Omega$/mW and $P_s^\text{CRI} = 25.7$~mW at the CRI side). The power dependence of the ratio $A^{\text{CRA}}/A^{\text{CRI}}$ of the fits is illustrated by a dashed curve in panel (c).
    }
		\label{FigR1_twin} 
	\end{figure*}

\begin{figure*}
		\centering \includegraphics[width=0.8\linewidth]{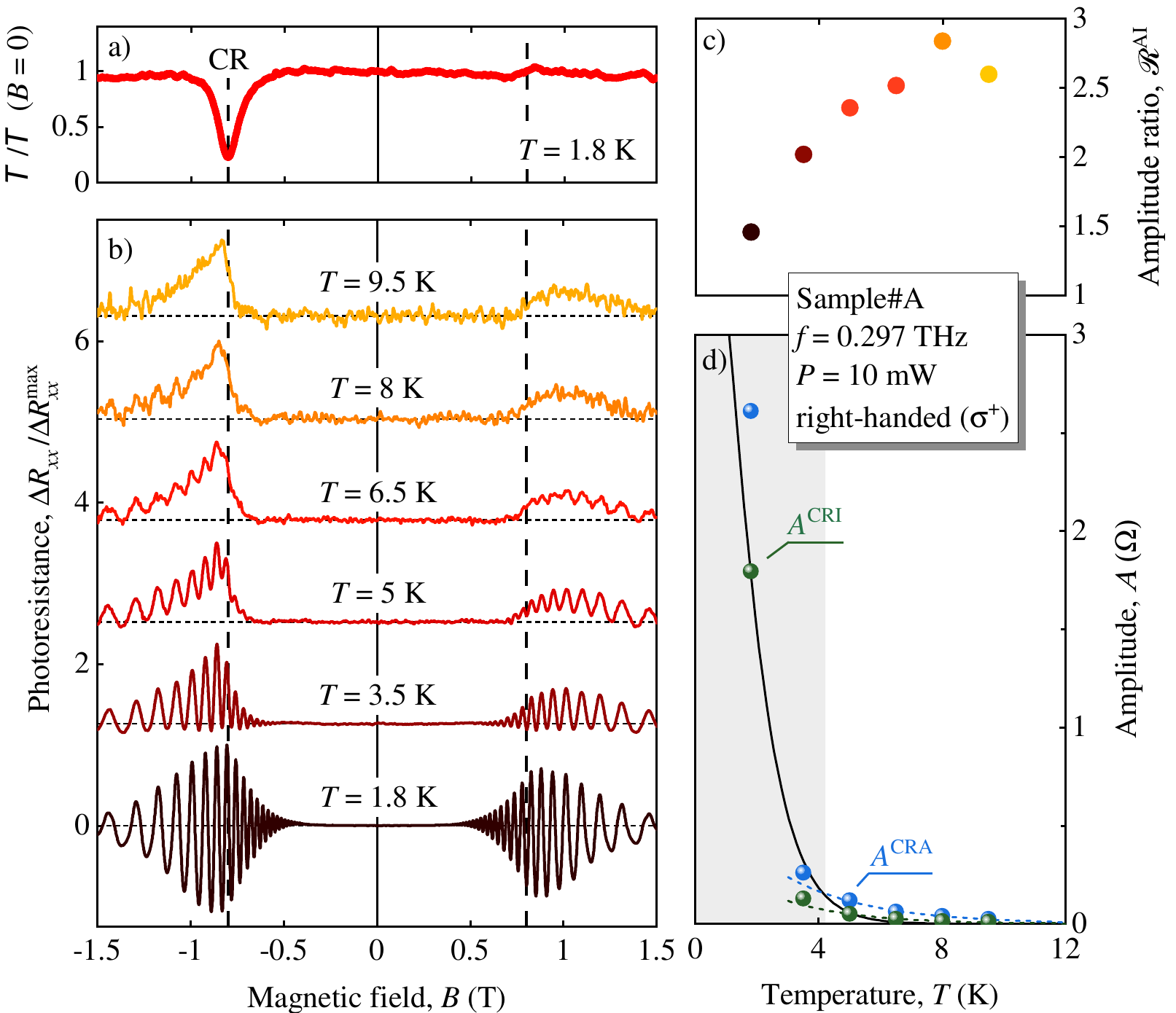}
		\caption{ Results obtained in sample \#A without prior exposure to room light under $P=10$~mW right-handed circularly polarized radiation ($\sigma^+$) produced by the IMPATT diode operating at frequency $f = 0.297$~THz. (a) Radiation transmittance normalized to its value at zero magnetic field, $\mathcal{T}/\mathcal{T}(B=0)$. (b) Normalized photoresistance traces at various temperatures. Vertical dashed lines mark the position of the CR for both helicities. (c) Temperature dependence of the amplitude ratio $\mathscr{R}^{\rm AI}= A^\text{CRA}/A^\text{CRI}$. (d) Temperature dependencies of the individual amplitudes $A^{\text{CRA}}$ and $A^{\text{CRI}}$. The grey shaded area marks the interval of $T \leq 4.2$~K where the SdHO-periodic oscillations in $\Delta R_{xx}$ remain strong. The corresponding solid line is calculated using the Lifshitz-Kosevich formula, see Sec.~\ref{discussion}.
        The dashed blue and green curves serve as guide for the eye, in the region where the SdHO are fully suppressed.
        }
		\label{FigR12} 
	\end{figure*}

\begin{figure*}
		\centering \includegraphics[width=0.8\linewidth]{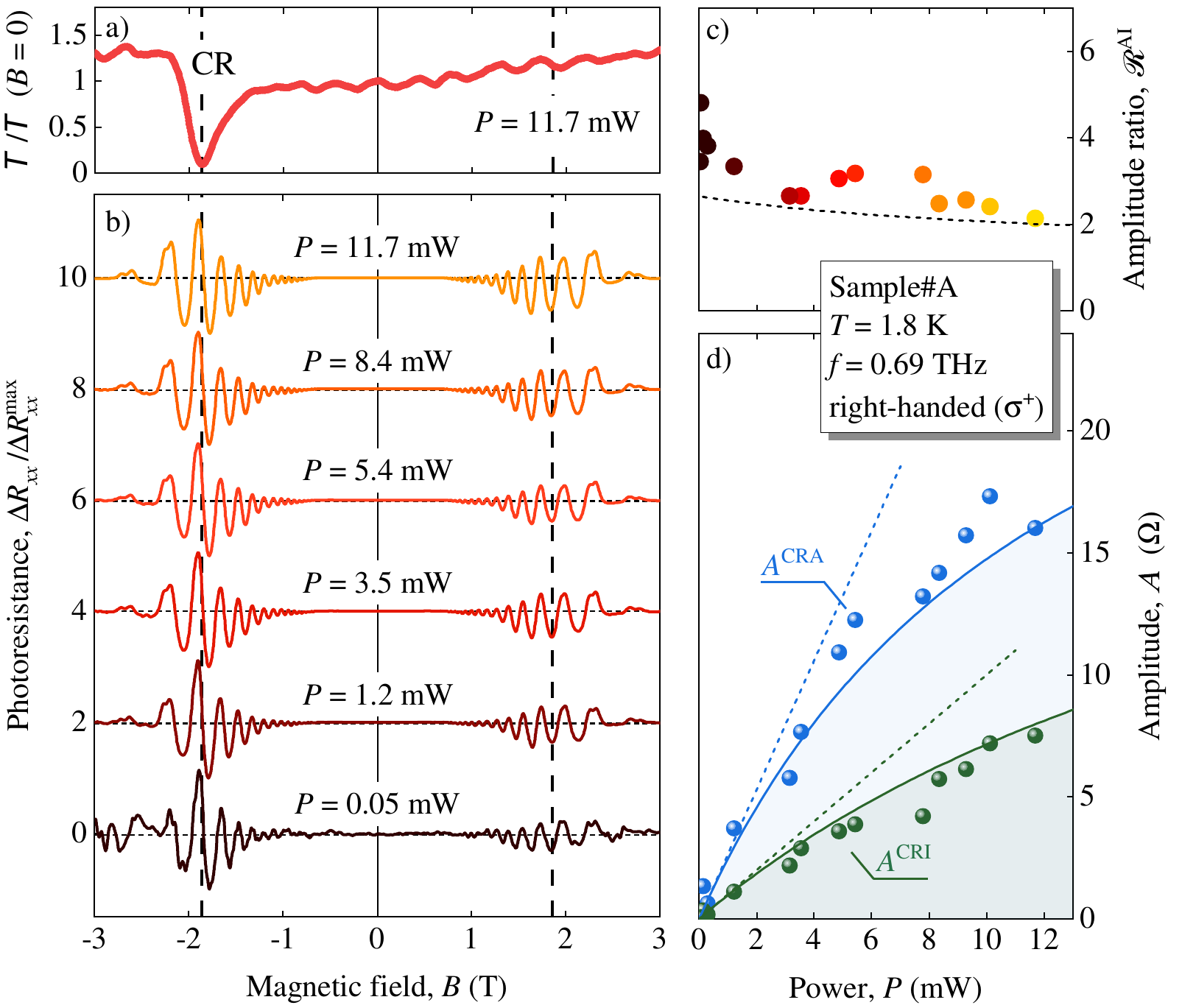}
		\caption{Results obtained in sample \#A at $T=1.8$~K without prior room light illumination and subjected to
  right-handed circularly polarized radiation ($\sigma^+$) produced by the molecular gas laser operating at frequency $f = 0.69$~THz. 
  (a) Radiation transmittance recorded at a radiation power of $P = 11.7$~mW and normalized to its value at zero magnetic field, $\mathcal{T}/\mathcal{T}(B=0)$. 
  (b) Photoresistance measured at various radiation powers. The traces are normalized to the signal's maximum, $\Delta R_{xx}/\Delta R_{xx}^{\text{max}}$ and are up-shifted for clarity. The position of the CR is indicated by vertical dashed lines for both helicities. (c) Power dependence of the ratio $\mathscr{R}^{\rm AI}= A^{\text{CRA}}/A^{\text{CRI}}$ of the CRA and CRI amplitudes. (d) Power dependencies of $A^{\text{CRA}}$ (blue) and $A^{\text{CRI}}$ (green), as extracted from unnormalized photoresistance data $\Delta R_{xx}$ measured at $f = 0.69$~THz. Solid lines are calculated according to Eq.~\eqref{saturation} using $a$ and $P_s$ as fitting parameters ($a^\text{CRA} = 2.65$~$\Omega$/mW and $P_s^\text{CRA} = 12.6$~mW at the CRA side, $a^\text{CRI} = 1.00$~$\Omega$/mW and $P_s^\text{CRI} = 25.1$~mW at the CRI side). Dashed lines illustrate the linear part $A=a P$ of the corresponding fits. The power dependence of the ratio $A^{\text{CRA}}/A^{\text{CRI}}$ of the fits is depicted by a dashed curve in panel (c).
  }
		\label{FigR16} 
	\end{figure*}

\begin{figure*}
		\centering \includegraphics[width=0.8\linewidth]{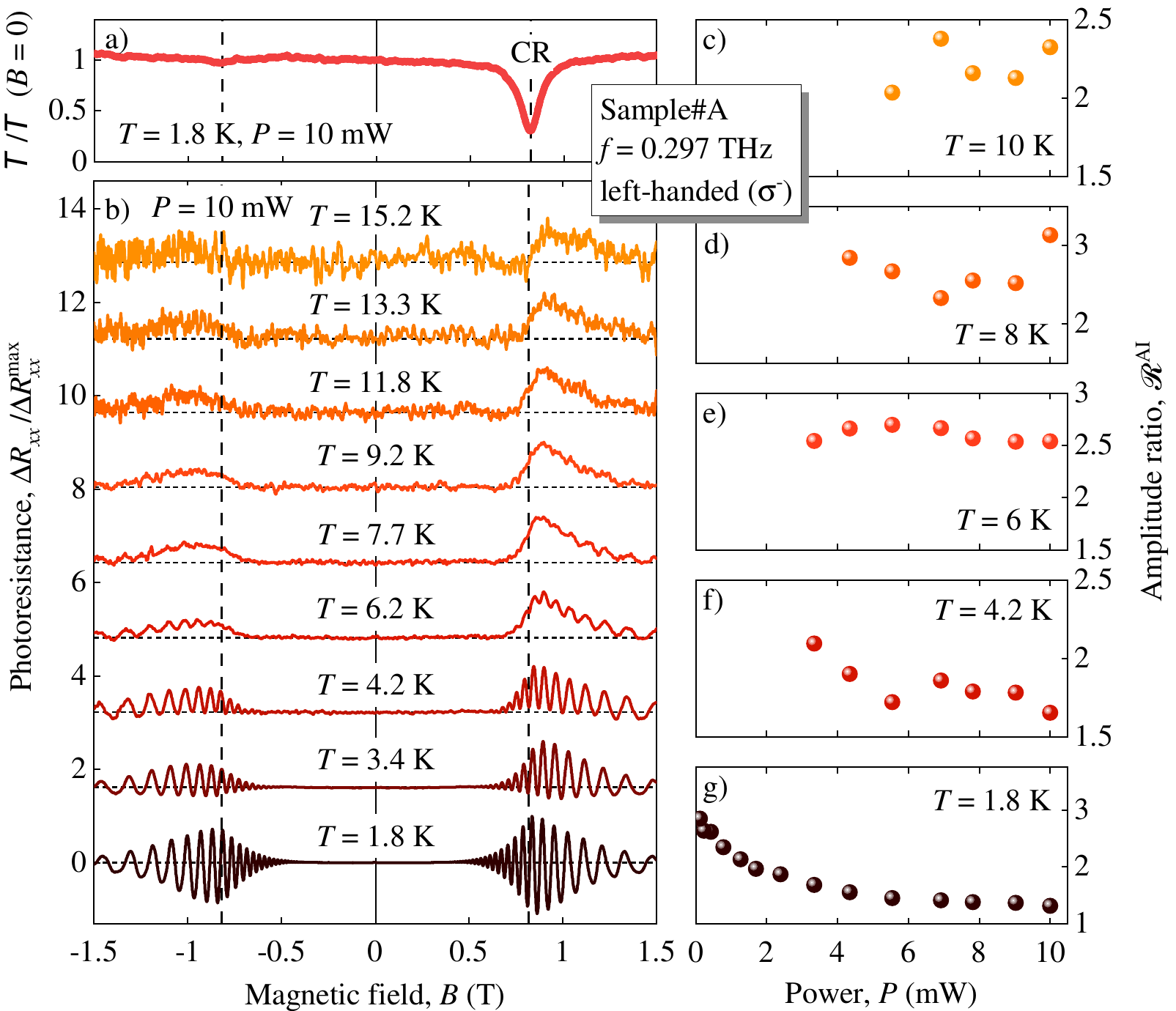}
		\caption{Results of combined temperature and power dependencies of photoresistance obtained on sample\#A at $f = 0.297$~THz and left-handed ($\sigma^-$) circularly polarized radiation. Sample was not illuminated by room light prior to measurements. (a) Normalized transmittance, $\mathcal{T}/\mathcal{T}(B=0)$, at temperature $T = 1.8$~K and power $P = 10$~mW. The CR positions for both helicities are indicated by vertical dashed lines. (b) The normalized photoresistance, $\Delta R_{xx}/\Delta R_{xx}^{\text{max}}$, at $P = 10$~mW for different temperatures. (c)- (g) The power dependencies of the CRA/CRI ratio $A^{\text{CRA}}/A^{\text{CRI}}$ extracted from measurements at given temperatures. 
  }
		\label{FigR13_3} 
	\end{figure*}

\begin{figure*}
		\centering \includegraphics[width=0.8\linewidth]{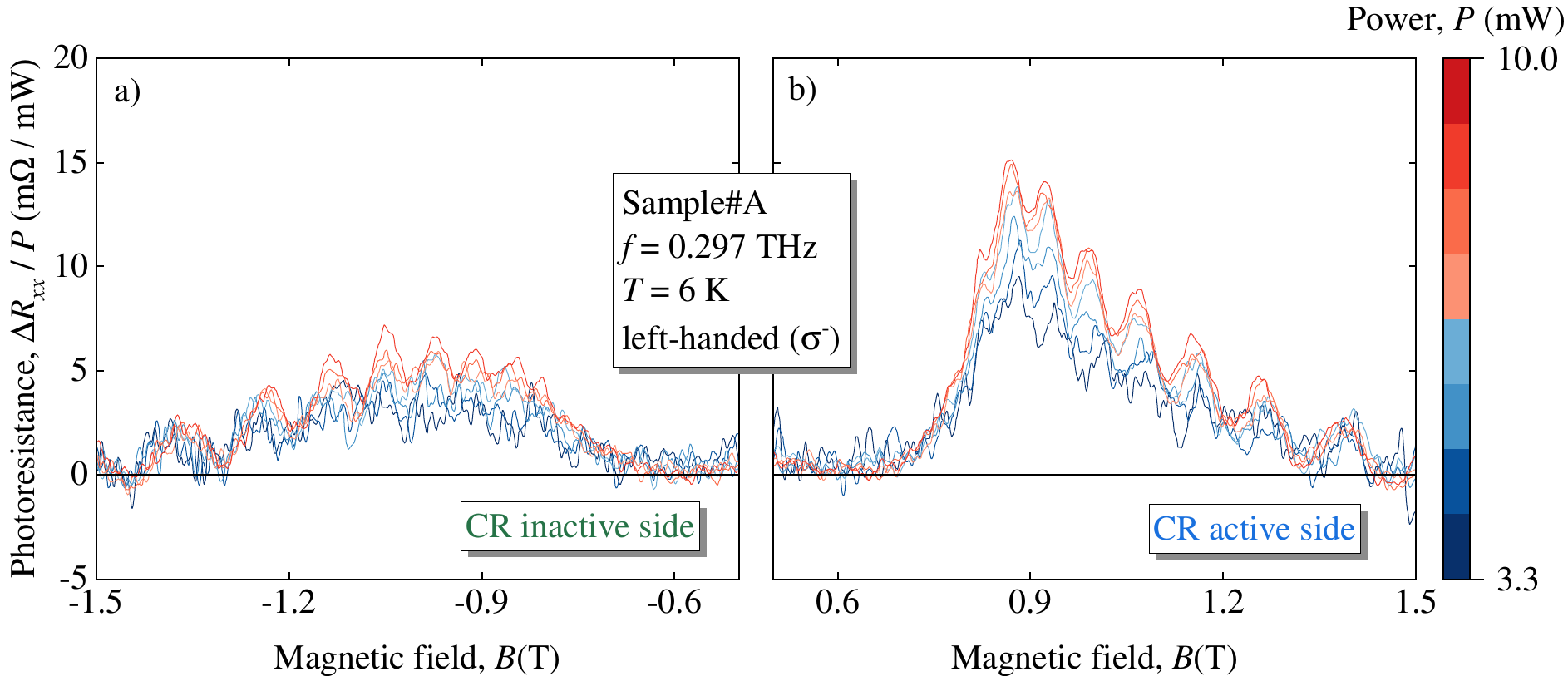}
		\caption{Power dependence of the photoresistance, $\Delta R_{xx}$, measured on sample \#A at $T = 6$~K and left-handed ($\sigma^-$) circularly polarized radiation of frequency $f = 0.297$~THz. The traces are normalized to their respective radiation power $P$. Panels (a) and (b) show the resonant photoresponse for CRI and CRA configurations, respectively. The colour of individual traces changes from intense blue at the lowest ($P=3.3$~mW) to intense red at the highest power ($P=10$~mW), as illustrated by the color bar on the right.
  }
		\label{FigR13bis} 
	\end{figure*}

\begin{figure*}
		\centering \includegraphics[width=0.8\linewidth]{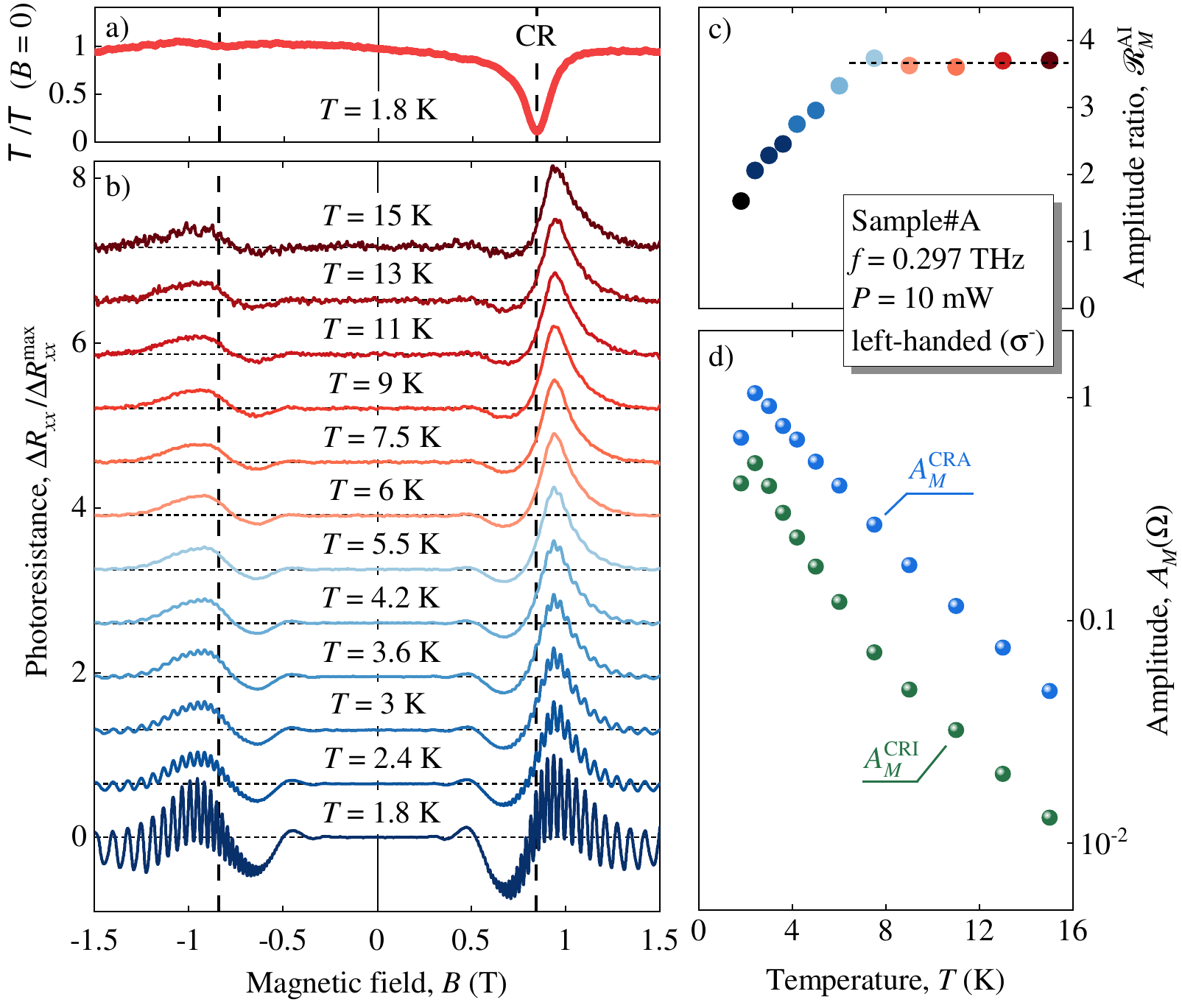}
		\caption{
    Results obtained on sample \#A after exposing it to room light and using left-handed circularly polarized radiation ($\sigma^-$) that was produced by the IMPATT diode with a frequency $f = 0.297$~THz and power $P = 10$~mW. (a) Radiation transmittance recorded at radiation power $P = 10$~mW and normalized to its value at zero magnetic field, $\mathcal{T}/\mathcal{T}(B=0)$. (b) Photoresistance measured at varying temperatures. The traces are normalized to their respective maximum, $\Delta R_{xx} / \Delta R_{xx}^\text{max}$, and are presented with an offset for clarity. The CR positions are marked by vertical dashed lines for both helicities. (c) Power dependence of the ratio $\mathscr{R}_M^\text{AI} = A_M^\text{CRA} / A_M^\text{CRI}$ of MIRO amplitudes, see Eq.~\eqref{ratioM}. (d) Power dependence of $A_M^\text{CRA}$ (blue) and $A_M^\text{CRI}$ (green).}
		\label{FigR3} 
	\end{figure*}

\begin{figure*}
		\centering \includegraphics[width=0.8\linewidth]{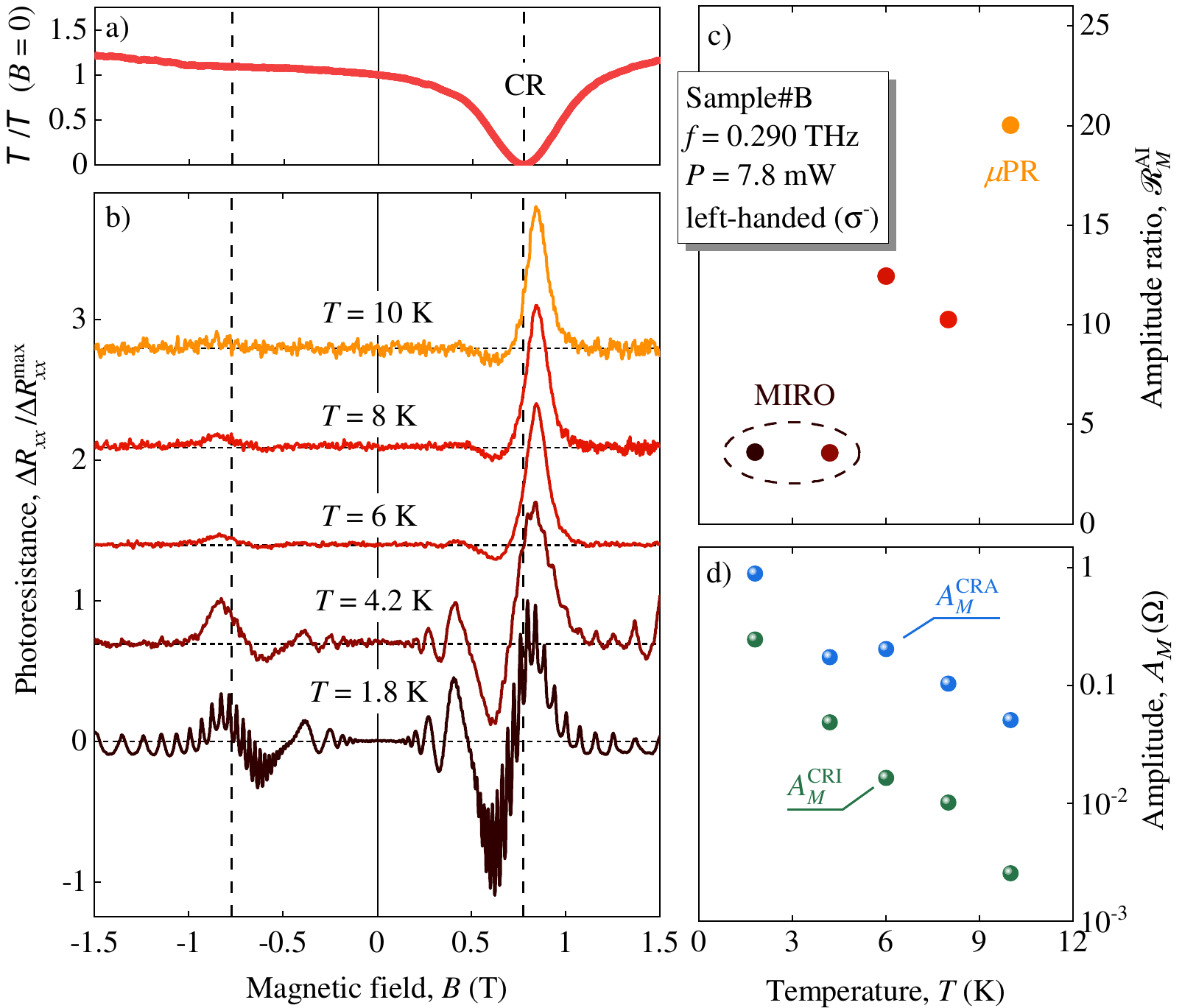}
		\caption{(a, b) Normalized transmittance and temperature dependence of the corresponding photoresistance measured at $P = 7.8$~mW and a frequency of $f = 0.290$~THz using the IMPATT diode setup. The data were obtained applying left-handed circularly polarized radiation ($\sigma^-$) to sample \#B, which was illuminated with ambient light prior to the measurements. Vertical dashed lines show the position of the CR for both helicities. (c) The CRA/CRI ratio $\mathscr{R}_M^{\rm AI} = A^{\text{CRA}}_M/A^{\text{CRI}}_M$, see Eq.~\eqref{ratioM}, as a function of temperature. The ellipse labels the points where MIRO is clearly visible in the photoresistance. The orange data point, marked by $\mu$PR, is dominated by the bolometric effect, here the $\mu$-photoconductivity. (d) The temperature dependence of individual signal amplitudes $A^{\rm CRA}_M$ and $A^{\rm CRI}_M$ plotted in a lin-log scale.
			}
		\label{FigR4} 
	\end{figure*}

In Figs.~\ref{FigR1_twin}-\ref{FigR4} we show the results and analysis of additional measurements which complement and support the data presented and discussed in the main text. 

Figure \ref{FigR1_twin} supports the results presented in Fig.~\ref{FigR1} of the main text. The data are obtained under the same conditions as there, but now for $\sigma^+$ circularly polarized radiation. It is seen that the whole behavior is closely reproduced for the opposite polarity, including the substantial growth of the CRA/CRI ratio at low radiation power.
Similarly, Figs.~\ref{FigR12} and \ref{FigR16} are the $\sigma^+$-analogues of Figs.~\ref{FigR11} and \ref{FigR5}, correspondingly, are also following the behavior detected for the opposite helicity. 

Figure \ref{FigR13_3} summarizes the studies of power dependencies of the photoresistance conducted at different $T$ on sample \#A. Apart from several selected photoresistance traces at $P=10$~mW and different $T$, here we present the power dependencies of the CRA/CRI ratio $\mathscr{R}_\text{AI}$ for various $T$ obtained from a large collection of such traces obtained at different $T$ and $P$. Figure \ref{FigR13bis} illustrates the power evolution of the shape of the photoresistance curves for $T = 6$~K by layering curves normalized by power to illustrate the shape progression. In contrast to $T=1.8$~K, see Fig.~\ref{FigR13}, it is seen that at such higher $T$, the nonlinear effects are much less pronounced, the shape remains almost the same at all $P$, and the $\Delta R_{xx}/P$ traces almost coincide for all $P$. Correspondingly, at high $T$ we do not observe any significant variations of the CRA/CRI ratio with the radiation power, see Fig.~\ref{FigR13_3}.  

Figure \ref{FigR3} shows the temperature dependence $\Delta R_{xx}$ for $\sigma^-$ circularly polarized radiation of sample \#A that was illuminated by room light before conducting the measurements, see also the power dependence at fixed low $T$ presented in Fig.~\ref{FigR2} and similar results obtained on sample \#B in Fig.~\ref{FigR17}. As in Fig.~\ref{FigR2}, here we also show the corresponding transmittance trace, as well as the analysis of the amplitudes $A_M^\text{CRA}$ and $A_M^\text{CRI}$ and their ratio $\mathscr{R}_M^\text{AI}$, as defined in Sec.~\ref{sample_A_room}. Figure \ref{FigR4} presents the transmittance and temperature dependence of the photoresistance measured at $f = 0.290$~THz on sample \#B. Similar to Fig.~\ref{FigR17}, the CRA/CRI ratio becomes substantially larger at high $T$, where the photoresistance is dominated by bolometric effects (marked $\mu$PR in the figure).

\bibliography{all_lib}

\begin{thebibliography}{51}%
\makeatletter
\providecommand \@ifxundefined [1]{%
 \@ifx{#1\undefined}
}%
\providecommand \@ifnum [1]{%
 \ifnum #1\expandafter \@firstoftwo
 \else \expandafter \@secondoftwo
 \fi
}%
\providecommand \@ifx [1]{%
 \ifx #1\expandafter \@firstoftwo
 \else \expandafter \@secondoftwo
 \fi
}%
\providecommand \natexlab [1]{#1}%
\providecommand \enquote  [1]{``#1''}%
\providecommand \bibnamefont  [1]{#1}%
\providecommand \bibfnamefont [1]{#1}%
\providecommand \citenamefont [1]{#1}%
\providecommand \href@noop [0]{\@secondoftwo}%
\providecommand \href [0]{\begingroup \@sanitize@url \@href}%
\providecommand \@href[1]{\@@startlink{#1}\@@href}%
\providecommand \@@href[1]{\endgroup#1\@@endlink}%
\providecommand \@sanitize@url [0]{\catcode `\\12\catcode `\$12\catcode
  `\&12\catcode `\#12\catcode `\^12\catcode `\_12\catcode `\%12\relax}%
\providecommand \@@startlink[1]{}%
\providecommand \@@endlink[0]{}%
\providecommand \url  [0]{\begingroup\@sanitize@url \@url }%
\providecommand \@url [1]{\endgroup\@href {#1}{\urlprefix }}%
\providecommand \urlprefix  [0]{URL }%
\providecommand \Eprint [0]{\href }%
\providecommand \doibase [0]{https://doi.org/}%
\providecommand \selectlanguage [0]{\@gobble}%
\providecommand \bibinfo  [0]{\@secondoftwo}%
\providecommand \bibfield  [0]{\@secondoftwo}%
\providecommand \translation [1]{[#1]}%
\providecommand \BibitemOpen [0]{}%
\providecommand \bibitemStop [0]{}%
\providecommand \bibitemNoStop [0]{.\EOS\space}%
\providecommand \EOS [0]{\spacefactor3000\relax}%
\providecommand \BibitemShut  [1]{\csname bibitem#1\endcsname}%
\let\auto@bib@innerbib\@empty
\bibitem [{\citenamefont {Seeger}(2004)}]{Seeger2004}%
  \BibitemOpen
  \bibfield  {author} {\bibinfo {author} {\bibfnamefont {K.}~\bibnamefont
  {Seeger}},\ }\href
  {https://doi.org/https://doi.org/10.1007/978-3-662-09855-4} {\emph {\bibinfo
  {title} {Semiconductor Physics: An Introduction}}}\ (\bibinfo  {publisher}
  {Springer},\ \bibinfo {year} {2004})\BibitemShut {NoStop}%
\bibitem [{\citenamefont {Hilton}\ \emph {et~al.}(2012)\citenamefont {Hilton},
  \citenamefont {Arikawa},\ and\ \citenamefont {Kono}}]{Hilton2012}%
  \BibitemOpen
  \bibfield  {author} {\bibinfo {author} {\bibfnamefont {D.~J.}\ \bibnamefont
  {Hilton}}, \bibinfo {author} {\bibfnamefont {T.}~\bibnamefont {Arikawa}},\
  and\ \bibinfo {author} {\bibfnamefont {J.}~\bibnamefont {Kono}},\ }\bibfield
  {title} {\bibinfo {title} {Cyclotron resonance},\ }\href@noop {} {\bibfield
  {journal} {\bibinfo  {journal} {Characterization of Materials, edited by
  Elton N. Kaufmann, (John Wiley \& Sons, NewYork)}\ ,\ \bibinfo {pages} {1}}
  (\bibinfo {year} {2012})}\BibitemShut {NoStop}%
\bibitem [{\citenamefont {Zudov}\ \emph {et~al.}(2001)\citenamefont {Zudov},
  \citenamefont {Du}, \citenamefont {Simmons},\ and\ \citenamefont
  {Reno}}]{Zudov2001}%
  \BibitemOpen
  \bibfield  {author} {\bibinfo {author} {\bibfnamefont {M.~A.}\ \bibnamefont
  {Zudov}}, \bibinfo {author} {\bibfnamefont {R.~R.}\ \bibnamefont {Du}},
  \bibinfo {author} {\bibfnamefont {J.~A.}\ \bibnamefont {Simmons}},\ and\
  \bibinfo {author} {\bibfnamefont {J.~L.}\ \bibnamefont {Reno}},\ }\bibfield
  {title} {\bibinfo {title} {Shubnikov{\textendash}de haas-like oscillations in
  millimeterwave photoconductivity in a high-mobility two-dimensional electron
  gas},\ }\href {https://doi.org/10.1103/physrevb.64.201311} {\bibfield
  {journal} {\bibinfo  {journal} {Phys. Rev. B}\ }\textbf {\bibinfo {volume}
  {64}},\ \bibinfo {pages} {201311(R)} (\bibinfo {year} {2001})}\BibitemShut
  {NoStop}%
\bibitem [{\citenamefont {Dmitriev}\ \emph {et~al.}(2012)\citenamefont
  {Dmitriev}, \citenamefont {Mirlin}, \citenamefont {Polyakov},\ and\
  \citenamefont {Zudov}}]{Dmitriev2012}%
  \BibitemOpen
  \bibfield  {author} {\bibinfo {author} {\bibfnamefont {I.~A.}\ \bibnamefont
  {Dmitriev}}, \bibinfo {author} {\bibfnamefont {A.~D.}\ \bibnamefont
  {Mirlin}}, \bibinfo {author} {\bibfnamefont {D.~G.}\ \bibnamefont
  {Polyakov}},\ and\ \bibinfo {author} {\bibfnamefont {M.~A.}\ \bibnamefont
  {Zudov}},\ }\bibfield  {title} {\bibinfo {title} {Nonequilibrium phenomena in
  high landau levels},\ }\href {https://doi.org/10.1103/revmodphys.84.1709}
  {\bibfield  {journal} {\bibinfo  {journal} {Rev. Mod. Phys.}\ }\textbf
  {\bibinfo {volume} {84}},\ \bibinfo {pages} {1709} (\bibinfo {year}
  {2012})}\BibitemShut {NoStop}%
\bibitem [{\citenamefont {Bandurin}\ \emph {et~al.}(2022)\citenamefont
  {Bandurin}, \citenamefont {Mönch}, \citenamefont {Kapralov}, \citenamefont
  {Phinney}, \citenamefont {Lindner}, \citenamefont {Liu}, \citenamefont
  {Edgar}, \citenamefont {Dmitriev}, \citenamefont {Jarillo-Herrero},
  \citenamefont {Svintsov},\ and\ \citenamefont {Ganichev}}]{Bandurin2022}%
  \BibitemOpen
  \bibfield  {author} {\bibinfo {author} {\bibfnamefont {D.~A.}\ \bibnamefont
  {Bandurin}}, \bibinfo {author} {\bibfnamefont {E.}~\bibnamefont {Mönch}},
  \bibinfo {author} {\bibfnamefont {K.}~\bibnamefont {Kapralov}}, \bibinfo
  {author} {\bibfnamefont {I.~Y.}\ \bibnamefont {Phinney}}, \bibinfo {author}
  {\bibfnamefont {K.}~\bibnamefont {Lindner}}, \bibinfo {author} {\bibfnamefont
  {S.}~\bibnamefont {Liu}}, \bibinfo {author} {\bibfnamefont {J.~H.}\
  \bibnamefont {Edgar}}, \bibinfo {author} {\bibfnamefont {I.~A.}\ \bibnamefont
  {Dmitriev}}, \bibinfo {author} {\bibfnamefont {P.}~\bibnamefont
  {Jarillo-Herrero}}, \bibinfo {author} {\bibfnamefont {D.}~\bibnamefont
  {Svintsov}},\ and\ \bibinfo {author} {\bibfnamefont {S.~D.}\ \bibnamefont
  {Ganichev}},\ }\bibfield  {title} {\bibinfo {title} {Cyclotron resonance
  overtones and near-field magnetoabsorption via terahertz bernstein modes in
  graphene},\ }\href {https://doi.org/10.1038/s41567-021-01494-8} {\bibfield
  {journal} {\bibinfo  {journal} {Nat. Phys.}\ }\textbf {\bibinfo {volume}
  {18}},\ \bibinfo {pages} {462–467} (\bibinfo {year} {2022})}\BibitemShut
  {NoStop}%
\bibitem [{\citenamefont {Dmitriev}\ \emph {et~al.}(2008)\citenamefont
  {Dmitriev}, \citenamefont {Evers}, \citenamefont {Gornyi}, \citenamefont
  {Mirlin}, \citenamefont {Polyakov},\ and\ \citenamefont
  {W\"{o}lfle}}]{Dmitriev2008}%
  \BibitemOpen
  \bibfield  {author} {\bibinfo {author} {\bibfnamefont {I.~A.}\ \bibnamefont
  {Dmitriev}}, \bibinfo {author} {\bibfnamefont {F.}~\bibnamefont {Evers}},
  \bibinfo {author} {\bibfnamefont {I.~V.}\ \bibnamefont {Gornyi}}, \bibinfo
  {author} {\bibfnamefont {A.~D.}\ \bibnamefont {Mirlin}}, \bibinfo {author}
  {\bibfnamefont {D.~G.}\ \bibnamefont {Polyakov}},\ and\ \bibinfo {author}
  {\bibfnamefont {P.}~\bibnamefont {W\"{o}lfle}},\ }\bibfield  {title}
  {\bibinfo {title} {Magnetotransport of electrons in quantum hall systems},\
  }\href {https://doi.org/10.1002/pssb.200743278} {\bibfield  {journal}
  {\bibinfo  {journal} {physica status solidi (b)}\ }\textbf {\bibinfo {volume}
  {245}},\ \bibinfo {pages} {239} (\bibinfo {year} {2008})}\BibitemShut
  {NoStop}%
\bibitem [{\citenamefont {Smet}\ \emph {et~al.}(2005)\citenamefont {Smet},
  \citenamefont {Gorshunov}, \citenamefont {Jiang}, \citenamefont {Pfeiffer},
  \citenamefont {West}, \citenamefont {Umansky}, \citenamefont {Dressel},
  \citenamefont {Meisels}, \citenamefont {Kuchar},\ and\ \citenamefont {von
  Klitzing}}]{Smet2005}%
  \BibitemOpen
  \bibfield  {author} {\bibinfo {author} {\bibfnamefont {J.~H.}\ \bibnamefont
  {Smet}}, \bibinfo {author} {\bibfnamefont {B.}~\bibnamefont {Gorshunov}},
  \bibinfo {author} {\bibfnamefont {C.}~\bibnamefont {Jiang}}, \bibinfo
  {author} {\bibfnamefont {L.}~\bibnamefont {Pfeiffer}}, \bibinfo {author}
  {\bibfnamefont {K.}~\bibnamefont {West}}, \bibinfo {author} {\bibfnamefont
  {V.}~\bibnamefont {Umansky}}, \bibinfo {author} {\bibfnamefont
  {M.}~\bibnamefont {Dressel}}, \bibinfo {author} {\bibfnamefont
  {R.}~\bibnamefont {Meisels}}, \bibinfo {author} {\bibfnamefont
  {F.}~\bibnamefont {Kuchar}},\ and\ \bibinfo {author} {\bibfnamefont
  {K.}~\bibnamefont {von Klitzing}},\ }\bibfield  {title} {\bibinfo {title}
  {Circular-polarization-dependent study of the microwave photoconductivity in
  a two-dimensional electron system},\ }\href
  {https://doi.org/10.1103/physrevlett.95.116804} {\bibfield  {journal}
  {\bibinfo  {journal} {Phys. Rev. Lett.}\ }\textbf {\bibinfo {volume} {95}},\
  \bibinfo {pages} {116804} (\bibinfo {year} {2005})}\BibitemShut {NoStop}%
\bibitem [{\citenamefont {Herrmann}\ \emph {et~al.}(2016)\citenamefont
  {Herrmann}, \citenamefont {Dmitriev}, \citenamefont {Kozlov}, \citenamefont
  {Schneider}, \citenamefont {Jentzsch}, \citenamefont {Kvon}, \citenamefont
  {Olbrich}, \citenamefont {Bel'kov}, \citenamefont {Bayer}, \citenamefont
  {Schuh}, \citenamefont {Bougeard}, \citenamefont {Kuczmik}, \citenamefont
  {Oltscher}, \citenamefont {Weiss},\ and\ \citenamefont
  {Ganichev}}]{Herrmann2016}%
  \BibitemOpen
  \bibfield  {author} {\bibinfo {author} {\bibfnamefont {T.}~\bibnamefont
  {Herrmann}}, \bibinfo {author} {\bibfnamefont {I.~A.}\ \bibnamefont
  {Dmitriev}}, \bibinfo {author} {\bibfnamefont {D.~A.}\ \bibnamefont
  {Kozlov}}, \bibinfo {author} {\bibfnamefont {M.}~\bibnamefont {Schneider}},
  \bibinfo {author} {\bibfnamefont {B.}~\bibnamefont {Jentzsch}}, \bibinfo
  {author} {\bibfnamefont {Z.~D.}\ \bibnamefont {Kvon}}, \bibinfo {author}
  {\bibfnamefont {P.}~\bibnamefont {Olbrich}}, \bibinfo {author} {\bibfnamefont
  {V.~V.}\ \bibnamefont {Bel'kov}}, \bibinfo {author} {\bibfnamefont
  {A.}~\bibnamefont {Bayer}}, \bibinfo {author} {\bibfnamefont
  {D.}~\bibnamefont {Schuh}}, \bibinfo {author} {\bibfnamefont
  {D.}~\bibnamefont {Bougeard}}, \bibinfo {author} {\bibfnamefont
  {T.}~\bibnamefont {Kuczmik}}, \bibinfo {author} {\bibfnamefont
  {M.}~\bibnamefont {Oltscher}}, \bibinfo {author} {\bibfnamefont
  {D.}~\bibnamefont {Weiss}},\ and\ \bibinfo {author} {\bibfnamefont {S.~D.}\
  \bibnamefont {Ganichev}},\ }\bibfield  {title} {\bibinfo {title} {Analog of
  microwave-induced resistance oscillations induced in {GaAs} heterostructures
  by terahertz radiation},\ }\href {https://doi.org/10.1103/physrevb.94.081301}
  {\bibfield  {journal} {\bibinfo  {journal} {Phys. Rev. B}\ }\textbf {\bibinfo
  {volume} {94}},\ \bibinfo {pages} {081301 (R)} (\bibinfo {year}
  {2016})}\BibitemShut {NoStop}%
\bibitem [{\citenamefont {Herrmann}\ \emph {et~al.}(2017)\citenamefont
  {Herrmann}, \citenamefont {Kvon}, \citenamefont {Dmitriev}, \citenamefont
  {Kozlov}, \citenamefont {Jentzsch}, \citenamefont {Schneider}, \citenamefont
  {Schell}, \citenamefont {Bel'kov}, \citenamefont {Bayer}, \citenamefont
  {Schuh}, \citenamefont {Bougeard}, \citenamefont {Kuczmik}, \citenamefont
  {Oltscher}, \citenamefont {Weiss},\ and\ \citenamefont
  {Ganichev}}]{Herrmann2017}%
  \BibitemOpen
  \bibfield  {author} {\bibinfo {author} {\bibfnamefont {T.}~\bibnamefont
  {Herrmann}}, \bibinfo {author} {\bibfnamefont {Z.~D.}\ \bibnamefont {Kvon}},
  \bibinfo {author} {\bibfnamefont {I.~A.}\ \bibnamefont {Dmitriev}}, \bibinfo
  {author} {\bibfnamefont {D.~A.}\ \bibnamefont {Kozlov}}, \bibinfo {author}
  {\bibfnamefont {B.}~\bibnamefont {Jentzsch}}, \bibinfo {author}
  {\bibfnamefont {M.}~\bibnamefont {Schneider}}, \bibinfo {author}
  {\bibfnamefont {L.}~\bibnamefont {Schell}}, \bibinfo {author} {\bibfnamefont
  {V.~V.}\ \bibnamefont {Bel'kov}}, \bibinfo {author} {\bibfnamefont
  {A.}~\bibnamefont {Bayer}}, \bibinfo {author} {\bibfnamefont
  {D.}~\bibnamefont {Schuh}}, \bibinfo {author} {\bibfnamefont
  {D.}~\bibnamefont {Bougeard}}, \bibinfo {author} {\bibfnamefont
  {T.}~\bibnamefont {Kuczmik}}, \bibinfo {author} {\bibfnamefont
  {M.}~\bibnamefont {Oltscher}}, \bibinfo {author} {\bibfnamefont
  {D.}~\bibnamefont {Weiss}},\ and\ \bibinfo {author} {\bibfnamefont {S.~D.}\
  \bibnamefont {Ganichev}},\ }\bibfield  {title} {\bibinfo {title}
  {Magnetoresistance oscillations induced by high-intensity terahertz
  radiation},\ }\href {https://doi.org/10.1103/physrevb.96.115449} {\bibfield
  {journal} {\bibinfo  {journal} {Phys. Rev. B}\ }\textbf {\bibinfo {volume}
  {96}},\ \bibinfo {pages} {115449} (\bibinfo {year} {2017})}\BibitemShut
  {NoStop}%
\bibitem [{\citenamefont {M\"onch}\ \emph {et~al.}(2022)\citenamefont
  {M\"onch}, \citenamefont {Euringer}, \citenamefont {H\"uttner}, \citenamefont
  {Dmitriev}, \citenamefont {Schuh}, \citenamefont {Marocko}, \citenamefont
  {Eroms}, \citenamefont {Bougeard}, \citenamefont {Weiss},\ and\ \citenamefont
  {Ganichev}}]{Moench2022b}%
  \BibitemOpen
  \bibfield  {author} {\bibinfo {author} {\bibfnamefont {E.}~\bibnamefont
  {M\"onch}}, \bibinfo {author} {\bibfnamefont {P.}~\bibnamefont {Euringer}},
  \bibinfo {author} {\bibfnamefont {G.-M.}\ \bibnamefont {H\"uttner}}, \bibinfo
  {author} {\bibfnamefont {I.~A.}\ \bibnamefont {Dmitriev}}, \bibinfo {author}
  {\bibfnamefont {D.}~\bibnamefont {Schuh}}, \bibinfo {author} {\bibfnamefont
  {M.}~\bibnamefont {Marocko}}, \bibinfo {author} {\bibfnamefont
  {J.}~\bibnamefont {Eroms}}, \bibinfo {author} {\bibfnamefont
  {D.}~\bibnamefont {Bougeard}}, \bibinfo {author} {\bibfnamefont
  {D.}~\bibnamefont {Weiss}},\ and\ \bibinfo {author} {\bibfnamefont {S.~D.}\
  \bibnamefont {Ganichev}},\ }\bibfield  {title} {\bibinfo {title} {Circular
  polarization immunity of the cyclotron resonance photoconductivity in
  two-dimensional electron systems},\ }\href
  {https://doi.org/10.1103/physrevb.106.l161409} {\bibfield  {journal}
  {\bibinfo  {journal} {Physical Review B}\ }\textbf {\bibinfo {volume}
  {106}},\ \bibinfo {pages} {L161409} (\bibinfo {year} {2022})}\BibitemShut
  {NoStop}%
\bibitem [{\citenamefont {Savchenko}\ \emph {et~al.}(2022)\citenamefont
  {Savchenko}, \citenamefont {Shuvaev}, \citenamefont {Dmitriev}, \citenamefont
  {Ganichev}, \citenamefont {Kvon},\ and\ \citenamefont
  {Pimenov}}]{Savchenko2022}%
  \BibitemOpen
  \bibfield  {author} {\bibinfo {author} {\bibfnamefont {M.~L.}\ \bibnamefont
  {Savchenko}}, \bibinfo {author} {\bibfnamefont {A.}~\bibnamefont {Shuvaev}},
  \bibinfo {author} {\bibfnamefont {I.~A.}\ \bibnamefont {Dmitriev}}, \bibinfo
  {author} {\bibfnamefont {S.~D.}\ \bibnamefont {Ganichev}}, \bibinfo {author}
  {\bibfnamefont {Z.~D.}\ \bibnamefont {Kvon}},\ and\ \bibinfo {author}
  {\bibfnamefont {A.}~\bibnamefont {Pimenov}},\ }\bibfield  {title} {\bibinfo
  {title} {Demonstration of high sensitivity of microwave-induced resistance
  oscillations to circular polarization},\ }\href
  {https://doi.org/10.1103/physrevb.106.l161408} {\bibfield  {journal}
  {\bibinfo  {journal} {Physical Review B}\ }\textbf {\bibinfo {volume}
  {106}},\ \bibinfo {pages} {L161408} (\bibinfo {year} {2022})}\BibitemShut
  {NoStop}%
\bibitem [{\citenamefont {Mooney}(1990)}]{Mooney1990}%
  \BibitemOpen
  \bibfield  {author} {\bibinfo {author} {\bibfnamefont {P.~M.}\ \bibnamefont
  {Mooney}},\ }\bibfield  {title} {\bibinfo {title} {Deep donor levels
  (\textit{DX} centers) in {III}-v semiconductors},\ }\href
  {https://doi.org/10.1063/1.345628} {\bibfield  {journal} {\bibinfo  {journal}
  {Journal of Appl. Phys.}\ }\textbf {\bibinfo {volume} {67}},\ \bibinfo
  {pages} {R1} (\bibinfo {year} {1990})}\BibitemShut {NoStop}%
\bibitem [{\citenamefont {Shur}(1990)}]{Shur1990}%
  \BibitemOpen
  \bibfield  {author} {\bibinfo {author} {\bibfnamefont {M.}~\bibnamefont
  {Shur}},\ }\href@noop {} {\emph {\bibinfo {title} {Physics of Semiconductor
  Devices}}}\ (\bibinfo  {publisher} {Prentice-Hall, Inc.},\ \bibinfo {address}
  {USA},\ \bibinfo {year} {1990})\BibitemShut {NoStop}%
\bibitem [{\citenamefont {Sze}(2008)}]{Sze2008}%
  \BibitemOpen
  \bibfield  {author} {\bibinfo {author} {\bibfnamefont {S.}~\bibnamefont
  {Sze}},\ }\href@noop {} {\emph {\bibinfo {title} {Semiconductor Devices:
  Physics and Technology, 2nd Ed.}}}\ (\bibinfo  {publisher} {Wiley India Pvt.
  Limited},\ \bibinfo {year} {2008})\BibitemShut {NoStop}%
\bibitem [{\citenamefont {Chantry}(1984)}]{Chantry1984}%
  \BibitemOpen
  \bibfield  {author} {\bibinfo {author} {\bibfnamefont {G.~W.}\ \bibnamefont
  {Chantry}},\ }\href@noop {} {\emph {\bibinfo {title} {Long-wave Optics : The
  Science and Technology of Infrared and Near-millimetre Waves}}}\ (\bibinfo
  {publisher} {Academic Press},\ \bibinfo {address} {London Orlando},\ \bibinfo
  {year} {1984})\BibitemShut {NoStop}%
\bibitem [{\citenamefont {Ganichev}\ and\ \citenamefont
  {Prettl}(2005)}]{Ganichev2005}%
  \BibitemOpen
  \bibfield  {author} {\bibinfo {author} {\bibfnamefont {S.~D.}\ \bibnamefont
  {Ganichev}}\ and\ \bibinfo {author} {\bibfnamefont {W.}~\bibnamefont
  {Prettl}},\ }\href
  {https://doi.org/10.1093/acprof:oso/9780198528302.001.0001} {\emph {\bibinfo
  {title} {Intense Terahertz Excitation of Semiconductors}}}\ (\bibinfo
  {publisher} {Oxford University Press},\ \bibinfo {address} {Oxford},\
  \bibinfo {year} {2005})\BibitemShut {NoStop}%
\bibitem [{\citenamefont {Bründermann}\ \emph {et~al.}(2012)\citenamefont
  {Bründermann}, \citenamefont {Hübers},\ and\ \citenamefont
  {Kimmitt}}]{Bruendermann2012}%
  \BibitemOpen
  \bibfield  {author} {\bibinfo {author} {\bibfnamefont {E.}~\bibnamefont
  {Bründermann}}, \bibinfo {author} {\bibfnamefont {H.-W.}\ \bibnamefont
  {Hübers}},\ and\ \bibinfo {author} {\bibfnamefont {M.~F.}\ \bibnamefont
  {Kimmitt}},\ }\href
  {https://www.ebook.de/de/product/9234425/erik_bruendermann_heinz_wilhelm_huebers_maurice_fitzgerald_kimmitt_terahertz_techniques.html}
  {\emph {\bibinfo {title} {Terahertz Techniques}}}\ (\bibinfo  {publisher}
  {Springer-Verlag GmbH},\ \bibinfo {year} {2012})\BibitemShut {NoStop}%
\bibitem [{\citenamefont {Kozlov}\ \emph {et~al.}(2011)\citenamefont {Kozlov},
  \citenamefont {Kvon}, \citenamefont {Mikhailov}, \citenamefont {Dvoretskii},\
  and\ \citenamefont {Portal}}]{Kozlov2011}%
  \BibitemOpen
  \bibfield  {author} {\bibinfo {author} {\bibfnamefont {D.}~\bibnamefont
  {Kozlov}}, \bibinfo {author} {\bibfnamefont {Z.~D.}\ \bibnamefont {Kvon}},
  \bibinfo {author} {\bibfnamefont {N.~N.}\ \bibnamefont {Mikhailov}}, \bibinfo
  {author} {\bibfnamefont {S.~A.}\ \bibnamefont {Dvoretskii}},\ and\ \bibinfo
  {author} {\bibfnamefont {J.~C.}\ \bibnamefont {Portal}},\ }\bibfield  {title}
  {\bibinfo {title} {Cyclotron resonance in a two.dimensional semimetal based
  on a hgte quantum well},\ }\href {https://doi.org/10.1134/S0021364011030088}
  {\bibfield  {journal} {\bibinfo  {journal} {JETP Lett.}\ }\textbf {\bibinfo
  {volume} {93}},\ \bibinfo {pages} {170} (\bibinfo {year} {2011})}\BibitemShut
  {NoStop}%
\bibitem [{\citenamefont {Otteneder}\ \emph {et~al.}(2018)\citenamefont
  {Otteneder}, \citenamefont {Dmitriev}, \citenamefont {Candussio},
  \citenamefont {Savchenko}, \citenamefont {Kozlov}, \citenamefont {Bel'kov},
  \citenamefont {Kvon}, \citenamefont {Mikhailov}, \citenamefont {Dvoretsky},\
  and\ \citenamefont {Ganichev}}]{Otteneder2018}%
  \BibitemOpen
  \bibfield  {author} {\bibinfo {author} {\bibfnamefont {M.}~\bibnamefont
  {Otteneder}}, \bibinfo {author} {\bibfnamefont {I.~A.}\ \bibnamefont
  {Dmitriev}}, \bibinfo {author} {\bibfnamefont {S.}~\bibnamefont {Candussio}},
  \bibinfo {author} {\bibfnamefont {M.~L.}\ \bibnamefont {Savchenko}}, \bibinfo
  {author} {\bibfnamefont {D.~A.}\ \bibnamefont {Kozlov}}, \bibinfo {author}
  {\bibfnamefont {V.~V.}\ \bibnamefont {Bel'kov}}, \bibinfo {author}
  {\bibfnamefont {Z.~D.}\ \bibnamefont {Kvon}}, \bibinfo {author}
  {\bibfnamefont {N.~N.}\ \bibnamefont {Mikhailov}}, \bibinfo {author}
  {\bibfnamefont {S.~A.}\ \bibnamefont {Dvoretsky}},\ and\ \bibinfo {author}
  {\bibfnamefont {S.~D.}\ \bibnamefont {Ganichev}},\ }\bibfield  {title}
  {\bibinfo {title} {Sign-alternating photoconductivity and magnetoresistance
  oscillations induced by terahertz radiation in hgte quantum wells},\ }\href
  {https://doi.org/10.1103/PhysRevB.98.245304} {\bibfield  {journal} {\bibinfo
  {journal} {Phys. Rev. B}\ }\textbf {\bibinfo {volume} {98}},\ \bibinfo
  {pages} {245304} (\bibinfo {year} {2018})}\BibitemShut {NoStop}%
\bibitem [{\citenamefont {Savchenko}\ \emph {et~al.}(2021)\citenamefont
  {Savchenko}, \citenamefont {Shuvaev}, \citenamefont {Dmitriev}, \citenamefont
  {Bykov}, \citenamefont {Bakarov}, \citenamefont {Kvon},\ and\ \citenamefont
  {Pimenov}}]{Savchenko2021}%
  \BibitemOpen
  \bibfield  {author} {\bibinfo {author} {\bibfnamefont {M.~L.}\ \bibnamefont
  {Savchenko}}, \bibinfo {author} {\bibfnamefont {A.}~\bibnamefont {Shuvaev}},
  \bibinfo {author} {\bibfnamefont {I.~A.}\ \bibnamefont {Dmitriev}}, \bibinfo
  {author} {\bibfnamefont {A.~A.}\ \bibnamefont {Bykov}}, \bibinfo {author}
  {\bibfnamefont {A.~K.}\ \bibnamefont {Bakarov}}, \bibinfo {author}
  {\bibfnamefont {Z.~D.}\ \bibnamefont {Kvon}},\ and\ \bibinfo {author}
  {\bibfnamefont {A.}~\bibnamefont {Pimenov}},\ }\bibfield  {title} {\bibinfo
  {title} {High harmonics of the cyclotron resonance in microwave transmission
  of a high-mobility two-dimensional electron system},\ }\href
  {https://doi.org/10.1103/physrevresearch.3.l012013} {\bibfield  {journal}
  {\bibinfo  {journal} {Phys. Rev. Research}\ }\textbf {\bibinfo {volume}
  {3}},\ \bibinfo {pages} {L012013} (\bibinfo {year} {2021})}\BibitemShut
  {NoStop}%
\bibitem [{\citenamefont {Hubmann}\ \emph {et~al.}(2019)\citenamefont
  {Hubmann}, \citenamefont {Gebert}, \citenamefont {Budkin}, \citenamefont
  {Bel'kov}, \citenamefont {Ivchenko}, \citenamefont {Dmitriev}, \citenamefont
  {Baumann}, \citenamefont {Otteneder}, \citenamefont {Ziegler}, \citenamefont
  {Disterheft}, \citenamefont {Kozlov}, \citenamefont {Mikhailov},
  \citenamefont {Dvoretsky}, \citenamefont {Kvon}, \citenamefont {Weiss},\ and\
  \citenamefont {Ganichev}}]{Hubmann2019}%
  \BibitemOpen
  \bibfield  {author} {\bibinfo {author} {\bibfnamefont {S.}~\bibnamefont
  {Hubmann}}, \bibinfo {author} {\bibfnamefont {S.}~\bibnamefont {Gebert}},
  \bibinfo {author} {\bibfnamefont {G.~V.}\ \bibnamefont {Budkin}}, \bibinfo
  {author} {\bibfnamefont {V.~V.}\ \bibnamefont {Bel'kov}}, \bibinfo {author}
  {\bibfnamefont {E.~L.}\ \bibnamefont {Ivchenko}}, \bibinfo {author}
  {\bibfnamefont {A.~P.}\ \bibnamefont {Dmitriev}}, \bibinfo {author}
  {\bibfnamefont {S.}~\bibnamefont {Baumann}}, \bibinfo {author} {\bibfnamefont
  {M.}~\bibnamefont {Otteneder}}, \bibinfo {author} {\bibfnamefont
  {J.}~\bibnamefont {Ziegler}}, \bibinfo {author} {\bibfnamefont
  {D.}~\bibnamefont {Disterheft}}, \bibinfo {author} {\bibfnamefont {D.~A.}\
  \bibnamefont {Kozlov}}, \bibinfo {author} {\bibfnamefont {N.~N.}\
  \bibnamefont {Mikhailov}}, \bibinfo {author} {\bibfnamefont {S.~A.}\
  \bibnamefont {Dvoretsky}}, \bibinfo {author} {\bibfnamefont {Z.~D.}\
  \bibnamefont {Kvon}}, \bibinfo {author} {\bibfnamefont {D.}~\bibnamefont
  {Weiss}},\ and\ \bibinfo {author} {\bibfnamefont {S.~D.}\ \bibnamefont
  {Ganichev}},\ }\bibfield  {title} {\bibinfo {title} {High-frequency impact
  ionization and nonlinearity of photocurrent induced by intense terahertz
  radiation in {HgTe}-based quantum well structures},\ }\href
  {https://doi.org/10.1103/physrevb.99.085312} {\bibfield  {journal} {\bibinfo
  {journal} {Phys. Rev. B}\ }\textbf {\bibinfo {volume} {99}},\ \bibinfo
  {pages} {085312} (\bibinfo {year} {2019})}\BibitemShut {NoStop}%
\bibitem [{\citenamefont {Candussio}\ \emph {et~al.}(2021)\citenamefont
  {Candussio}, \citenamefont {Golub}, \citenamefont {Bernreuter}, \citenamefont
  {J{\"o}tten}, \citenamefont {Rockinger}, \citenamefont {Watanabe},
  \citenamefont {Taniguchi}, \citenamefont {Eroms}, \citenamefont {Weiss},\
  and\ \citenamefont {Ganichev}}]{Candussio2021a}%
  \BibitemOpen
  \bibfield  {author} {\bibinfo {author} {\bibfnamefont {S.}~\bibnamefont
  {Candussio}}, \bibinfo {author} {\bibfnamefont {L.~E.}\ \bibnamefont
  {Golub}}, \bibinfo {author} {\bibfnamefont {S.}~\bibnamefont {Bernreuter}},
  \bibinfo {author} {\bibfnamefont {T.}~\bibnamefont {J{\"o}tten}}, \bibinfo
  {author} {\bibfnamefont {T.}~\bibnamefont {Rockinger}}, \bibinfo {author}
  {\bibfnamefont {K.}~\bibnamefont {Watanabe}}, \bibinfo {author}
  {\bibfnamefont {T.}~\bibnamefont {Taniguchi}}, \bibinfo {author}
  {\bibfnamefont {J.}~\bibnamefont {Eroms}}, \bibinfo {author} {\bibfnamefont
  {D.}~\bibnamefont {Weiss}},\ and\ \bibinfo {author} {\bibfnamefont {S.~D.}\
  \bibnamefont {Ganichev}},\ }\bibfield  {title} {\bibinfo {title} {Nonlinear
  intensity dependence of edge photocurrents in graphene induced by terahertz
  radiation},\ }\href {https://doi.org/10.1103/physrevb.104.155404} {\bibfield
  {journal} {\bibinfo  {journal} {Phys. Rev. B}\ }\textbf {\bibinfo {volume}
  {104}},\ \bibinfo {pages} {155404} (\bibinfo {year} {2021})}\BibitemShut
  {NoStop}%
\bibitem [{\citenamefont {Abstreiter}\ \emph {et~al.}(1976)\citenamefont
  {Abstreiter}, \citenamefont {Kotthaus}, \citenamefont {Koch},\ and\
  \citenamefont {Dorda}}]{Abstreiter1976}%
  \BibitemOpen
  \bibfield  {author} {\bibinfo {author} {\bibfnamefont {G.}~\bibnamefont
  {Abstreiter}}, \bibinfo {author} {\bibfnamefont {J.~P.}\ \bibnamefont
  {Kotthaus}}, \bibinfo {author} {\bibfnamefont {J.~F.}\ \bibnamefont {Koch}},\
  and\ \bibinfo {author} {\bibfnamefont {G.}~\bibnamefont {Dorda}},\ }\bibfield
   {title} {\bibinfo {title} {Cyclotron resonance of electrons in surface
  space-charge layers on silicon},\ }\href
  {https://doi.org/10.1103/physrevb.14.2480} {\bibfield  {journal} {\bibinfo
  {journal} {Phys. Rev. B}\ }\textbf {\bibinfo {volume} {14}},\ \bibinfo
  {pages} {2480–2493} (\bibinfo {year} {1976})}\BibitemShut {NoStop}%
\bibitem [{\citenamefont {Chiu}\ \emph {et~al.}(1976)\citenamefont {Chiu},
  \citenamefont {Lee},\ and\ \citenamefont {Quinn}}]{Chiu1976}%
  \BibitemOpen
  \bibfield  {author} {\bibinfo {author} {\bibfnamefont {K.}~\bibnamefont
  {Chiu}}, \bibinfo {author} {\bibfnamefont {T.}~\bibnamefont {Lee}},\ and\
  \bibinfo {author} {\bibfnamefont {J.}~\bibnamefont {Quinn}},\ }\bibfield
  {title} {\bibinfo {title} {Infrared magneto-transmittance of a
  two-dimensional electron gas},\ }\href
  {https://doi.org/https://doi.org/10.1016/0039-6028(76)90132-1} {\bibfield
  {journal} {\bibinfo  {journal} {Surf. Sci.}\ }\textbf {\bibinfo {volume}
  {58}},\ \bibinfo {pages} {182} (\bibinfo {year} {1976})}\BibitemShut
  {NoStop}%
\bibitem [{\citenamefont {Mikhailov}(2004)}]{Mikhailov2004}%
  \BibitemOpen
  \bibfield  {author} {\bibinfo {author} {\bibfnamefont {S.~A.}\ \bibnamefont
  {Mikhailov}},\ }\bibfield  {title} {\bibinfo {title} {Microwave-induced
  magnetotransport phenomena in two-dimensional electron systems: Importance of
  electrodynamic effects},\ }\href {https://doi.org/10.1103/PhysRevB.70.165311}
  {\bibfield  {journal} {\bibinfo  {journal} {Phys. Rev. B}\ }\textbf {\bibinfo
  {volume} {70}},\ \bibinfo {pages} {165311} (\bibinfo {year}
  {2004})}\BibitemShut {NoStop}%
\bibitem [{\citenamefont {Zhang}\ \emph {et~al.}(2014)\citenamefont {Zhang},
  \citenamefont {Arikawa}, \citenamefont {Kato}, \citenamefont {Reno},
  \citenamefont {Pan}, \citenamefont {Watson}, \citenamefont {Manfra},
  \citenamefont {Zudov}, \citenamefont {Tokman}, \citenamefont {Erukhimova},
  \citenamefont {Belyanin},\ and\ \citenamefont {Kono}}]{Zhang2014}%
  \BibitemOpen
  \bibfield  {author} {\bibinfo {author} {\bibfnamefont {Q.}~\bibnamefont
  {Zhang}}, \bibinfo {author} {\bibfnamefont {T.}~\bibnamefont {Arikawa}},
  \bibinfo {author} {\bibfnamefont {E.}~\bibnamefont {Kato}}, \bibinfo {author}
  {\bibfnamefont {J.~L.}\ \bibnamefont {Reno}}, \bibinfo {author}
  {\bibfnamefont {W.}~\bibnamefont {Pan}}, \bibinfo {author} {\bibfnamefont
  {J.~D.}\ \bibnamefont {Watson}}, \bibinfo {author} {\bibfnamefont {M.~J.}\
  \bibnamefont {Manfra}}, \bibinfo {author} {\bibfnamefont {M.~A.}\
  \bibnamefont {Zudov}}, \bibinfo {author} {\bibfnamefont {M.}~\bibnamefont
  {Tokman}}, \bibinfo {author} {\bibfnamefont {M.}~\bibnamefont {Erukhimova}},
  \bibinfo {author} {\bibfnamefont {A.}~\bibnamefont {Belyanin}},\ and\
  \bibinfo {author} {\bibfnamefont {J.}~\bibnamefont {Kono}},\ }\bibfield
  {title} {\bibinfo {title} {Superradiant decay of cyclotron resonance of
  two-dimensional electron gases},\ }\href
  {https://doi.org/10.1103/PhysRevLett.113.047601} {\bibfield  {journal}
  {\bibinfo  {journal} {Phys. Rev. Lett.}\ }\textbf {\bibinfo {volume} {113}},\
  \bibinfo {pages} {047601} (\bibinfo {year} {2014})}\BibitemShut {NoStop}%
\bibitem [{\citenamefont {Gantmakher}\ \emph {et~al.}(1988)\citenamefont
  {Gantmakher}, \citenamefont {Levinson}, \citenamefont {Grinberg},\ and\
  \citenamefont {Luryi}}]{Gantmakher1988}%
  \BibitemOpen
  \bibfield  {author} {\bibinfo {author} {\bibfnamefont {V.~F.}\ \bibnamefont
  {Gantmakher}}, \bibinfo {author} {\bibfnamefont {Y.~B.}\ \bibnamefont
  {Levinson}}, \bibinfo {author} {\bibfnamefont {A.~A.}\ \bibnamefont
  {Grinberg}},\ and\ \bibinfo {author} {\bibfnamefont {S.}~\bibnamefont
  {Luryi}},\ }\bibfield  {title} {\bibinfo {title} {Carrier scattering in
  metals and semiconductors},\ }\href {https://doi.org/10.1063/1.2811285}
  {\bibfield  {journal} {\bibinfo  {journal} {Phys. Today}\ }\textbf {\bibinfo
  {volume} {41}},\ \bibinfo {pages} {84} (\bibinfo {year} {1988})}\BibitemShut
  {NoStop}%
\bibitem [{\citenamefont {Muraviev}\ \emph {et~al.}(2013)\citenamefont
  {Muraviev}, \citenamefont {Rumyantsev}, \citenamefont {Liu}, \citenamefont
  {Balandin}, \citenamefont {Knap},\ and\ \citenamefont {Shur}}]{Muraviev2013}%
  \BibitemOpen
  \bibfield  {author} {\bibinfo {author} {\bibfnamefont {A.~V.}\ \bibnamefont
  {Muraviev}}, \bibinfo {author} {\bibfnamefont {S.~L.}\ \bibnamefont
  {Rumyantsev}}, \bibinfo {author} {\bibfnamefont {G.}~\bibnamefont {Liu}},
  \bibinfo {author} {\bibfnamefont {A.~A.}\ \bibnamefont {Balandin}}, \bibinfo
  {author} {\bibfnamefont {W.}~\bibnamefont {Knap}},\ and\ \bibinfo {author}
  {\bibfnamefont {M.~S.}\ \bibnamefont {Shur}},\ }\bibfield  {title} {\bibinfo
  {title} {Plasmonic and bolometric terahertz detection by graphene
  field-effect transistor},\ }\href {https://doi.org/10.1063/1.4826139}
  {\bibfield  {journal} {\bibinfo  {journal} {Appl. Phys. Lett.}\ }\textbf
  {\bibinfo {volume} {103}},\ \bibinfo {pages} {181114} (\bibinfo {year}
  {2013})}\BibitemShut {NoStop}%
\bibitem [{\citenamefont {Freitag}\ \emph {et~al.}(2012)\citenamefont
  {Freitag}, \citenamefont {Low}, \citenamefont {Xia},\ and\ \citenamefont
  {Avouris}}]{Freitag2012}%
  \BibitemOpen
  \bibfield  {author} {\bibinfo {author} {\bibfnamefont {M.}~\bibnamefont
  {Freitag}}, \bibinfo {author} {\bibfnamefont {T.}~\bibnamefont {Low}},
  \bibinfo {author} {\bibfnamefont {F.}~\bibnamefont {Xia}},\ and\ \bibinfo
  {author} {\bibfnamefont {P.}~\bibnamefont {Avouris}},\ }\bibfield  {title}
  {\bibinfo {title} {Photoconductivity of biased graphene},\ }\href
  {https://doi.org/10.1038/nphoton.2012.314} {\bibfield  {journal} {\bibinfo
  {journal} {Nat. Photonics}\ }\textbf {\bibinfo {volume} {7}},\ \bibinfo
  {pages} {53} (\bibinfo {year} {2012})}\BibitemShut {NoStop}%
\bibitem [{\citenamefont {Betz}\ \emph {et~al.}(2012)\citenamefont {Betz},
  \citenamefont {Vialla}, \citenamefont {Brunel}, \citenamefont {Voisin},
  \citenamefont {Picher}, \citenamefont {Cavanna}, \citenamefont {Madouri},
  \citenamefont {F{\`{e}}ve}, \citenamefont {Berroir}, \citenamefont
  {Pla{\c{c}}ais},\ and\ \citenamefont {Pallecchi}}]{Betz2012}%
  \BibitemOpen
  \bibfield  {author} {\bibinfo {author} {\bibfnamefont {A.~C.}\ \bibnamefont
  {Betz}}, \bibinfo {author} {\bibfnamefont {F.}~\bibnamefont {Vialla}},
  \bibinfo {author} {\bibfnamefont {D.}~\bibnamefont {Brunel}}, \bibinfo
  {author} {\bibfnamefont {C.}~\bibnamefont {Voisin}}, \bibinfo {author}
  {\bibfnamefont {M.}~\bibnamefont {Picher}}, \bibinfo {author} {\bibfnamefont
  {A.}~\bibnamefont {Cavanna}}, \bibinfo {author} {\bibfnamefont
  {A.}~\bibnamefont {Madouri}}, \bibinfo {author} {\bibfnamefont
  {G.}~\bibnamefont {F{\`{e}}ve}}, \bibinfo {author} {\bibfnamefont {J.-M.}\
  \bibnamefont {Berroir}}, \bibinfo {author} {\bibfnamefont {B.}~\bibnamefont
  {Pla{\c{c}}ais}},\ and\ \bibinfo {author} {\bibfnamefont {E.}~\bibnamefont
  {Pallecchi}},\ }\bibfield  {title} {\bibinfo {title} {Hot electron cooling by
  acoustic phonons in graphene},\ }\href
  {https://doi.org/10.1103/physrevlett.109.056805} {\bibfield  {journal}
  {\bibinfo  {journal} {Physical Review Letters}\ }\textbf {\bibinfo {volume}
  {109}},\ \bibinfo {pages} {056805} (\bibinfo {year} {2012})}\BibitemShut
  {NoStop}%
\bibitem [{\citenamefont {Heyman}\ \emph {et~al.}(2015)\citenamefont {Heyman},
  \citenamefont {Stein}, \citenamefont {Kaminski}, \citenamefont {Banman},
  \citenamefont {Massari},\ and\ \citenamefont {Robinson}}]{Heyman2015}%
  \BibitemOpen
  \bibfield  {author} {\bibinfo {author} {\bibfnamefont {J.~N.}\ \bibnamefont
  {Heyman}}, \bibinfo {author} {\bibfnamefont {J.~D.}\ \bibnamefont {Stein}},
  \bibinfo {author} {\bibfnamefont {Z.~S.}\ \bibnamefont {Kaminski}}, \bibinfo
  {author} {\bibfnamefont {A.~R.}\ \bibnamefont {Banman}}, \bibinfo {author}
  {\bibfnamefont {A.~M.}\ \bibnamefont {Massari}},\ and\ \bibinfo {author}
  {\bibfnamefont {J.~T.}\ \bibnamefont {Robinson}},\ }\bibfield  {title}
  {\bibinfo {title} {Carrier heating and negative photoconductivity in
  graphene},\ }\href {https://doi.org/10.1063/1.4905192} {\bibfield  {journal}
  {\bibinfo  {journal} {Journal of Applied Physics}\ }\textbf {\bibinfo
  {volume} {117}},\ \bibinfo {pages} {015101} (\bibinfo {year}
  {2015})}\BibitemShut {NoStop}%
\bibitem [{\citenamefont {Ryzhii}\ \emph {et~al.}(2019)\citenamefont {Ryzhii},
  \citenamefont {Ponomarev}, \citenamefont {Ryzhii}, \citenamefont {Mitin},
  \citenamefont {Shur},\ and\ \citenamefont {Otsuji}}]{Ryzhii2019}%
  \BibitemOpen
  \bibfield  {author} {\bibinfo {author} {\bibfnamefont {V.}~\bibnamefont
  {Ryzhii}}, \bibinfo {author} {\bibfnamefont {D.~S.}\ \bibnamefont
  {Ponomarev}}, \bibinfo {author} {\bibfnamefont {M.}~\bibnamefont {Ryzhii}},
  \bibinfo {author} {\bibfnamefont {V.}~\bibnamefont {Mitin}}, \bibinfo
  {author} {\bibfnamefont {M.~S.}\ \bibnamefont {Shur}},\ and\ \bibinfo
  {author} {\bibfnamefont {T.}~\bibnamefont {Otsuji}},\ }\bibfield  {title}
  {\bibinfo {title} {Negative and positive terahertz and infrared
  photoconductivity in uncooled graphene},\ }\href
  {https://doi.org/10.1364/ome.9.000585} {\bibfield  {journal} {\bibinfo
  {journal} {Optical Materials Express}\ }\textbf {\bibinfo {volume} {9}},\
  \bibinfo {pages} {585} (\bibinfo {year} {2019})}\BibitemShut {NoStop}%
\bibitem [{\citenamefont {Jago}\ \emph {et~al.}(2019)\citenamefont {Jago},
  \citenamefont {Malic},\ and\ \citenamefont {Wendler}}]{Jago2019}%
  \BibitemOpen
  \bibfield  {author} {\bibinfo {author} {\bibfnamefont {R.}~\bibnamefont
  {Jago}}, \bibinfo {author} {\bibfnamefont {E.}~\bibnamefont {Malic}},\ and\
  \bibinfo {author} {\bibfnamefont {F.}~\bibnamefont {Wendler}},\ }\bibfield
  {title} {\bibinfo {title} {Microscopic origin of the bolometric effect in
  graphene},\ }\href {https://doi.org/10.1103/physrevb.99.035419} {\bibfield
  {journal} {\bibinfo  {journal} {Physical Review B}\ }\textbf {\bibinfo
  {volume} {99}},\ \bibinfo {pages} {035419} (\bibinfo {year}
  {2019})}\BibitemShut {NoStop}%
\bibitem [{\citenamefont {Shoenberg}(1984)}]{Shoenberg1984}%
  \BibitemOpen
  \bibfield  {author} {\bibinfo {author} {\bibfnamefont {D.}~\bibnamefont
  {Shoenberg}},\ }\href {https://doi.org/10.1017/cbo9780511897870} {\emph
  {\bibinfo {title} {Magnetic Oscillations in Metals}}}\ (\bibinfo  {publisher}
  {Cambridge University Press},\ \bibinfo {year} {1984})\BibitemShut {NoStop}%
\bibitem [{\citenamefont {Ando}\ \emph {et~al.}(1982)\citenamefont {Ando},
  \citenamefont {Fowler},\ and\ \citenamefont {Stern}}]{Ando1982}%
  \BibitemOpen
  \bibfield  {author} {\bibinfo {author} {\bibfnamefont {T.}~\bibnamefont
  {Ando}}, \bibinfo {author} {\bibfnamefont {A.~B.}\ \bibnamefont {Fowler}},\
  and\ \bibinfo {author} {\bibfnamefont {F.}~\bibnamefont {Stern}},\ }\bibfield
   {title} {\bibinfo {title} {Electronic properties of two-dimensional
  systems},\ }\href {https://doi.org/10.1103/revmodphys.54.437} {\bibfield
  {journal} {\bibinfo  {journal} {Rev. Mod. Phys.}\ }\textbf {\bibinfo {volume}
  {54}},\ \bibinfo {pages} {437} (\bibinfo {year} {1982})}\BibitemShut
  {NoStop}%
\bibitem [{\citenamefont {Helm}\ \emph {et~al.}(1985)\citenamefont {Helm},
  \citenamefont {Gornik}, \citenamefont {Black}, \citenamefont {Allan},
  \citenamefont {Pidgeon}, \citenamefont {Mitchell},\ and\ \citenamefont
  {Weimann}}]{Helm1985}%
  \BibitemOpen
  \bibfield  {author} {\bibinfo {author} {\bibfnamefont {M.}~\bibnamefont
  {Helm}}, \bibinfo {author} {\bibfnamefont {E.}~\bibnamefont {Gornik}},
  \bibinfo {author} {\bibfnamefont {A.}~\bibnamefont {Black}}, \bibinfo
  {author} {\bibfnamefont {G.}~\bibnamefont {Allan}}, \bibinfo {author}
  {\bibfnamefont {C.}~\bibnamefont {Pidgeon}}, \bibinfo {author} {\bibfnamefont
  {K.}~\bibnamefont {Mitchell}},\ and\ \bibinfo {author} {\bibfnamefont
  {G.}~\bibnamefont {Weimann}},\ }\bibfield  {title} {\bibinfo {title} {Hot
  electron landau level lifetime in {GaAs}/{GaAlAs} heterostructures},\ }\href
  {https://doi.org/10.1016/0378-4363(85)90364-x} {\bibfield  {journal}
  {\bibinfo  {journal} {Physica B+C}\ }\textbf {\bibinfo {volume} {134}},\
  \bibinfo {pages} {323} (\bibinfo {year} {1985})}\BibitemShut {NoStop}%
\bibitem [{\citenamefont {Rodr{\'{\i}}guez}\ \emph {et~al.}(1986)\citenamefont
  {Rodr{\'{\i}}guez}, \citenamefont {Hart}, \citenamefont {Sievers},
  \citenamefont {Keilmann}, \citenamefont {Schlesinger}, \citenamefont
  {Wright},\ and\ \citenamefont {Wang}}]{Rodriguez1986}%
  \BibitemOpen
  \bibfield  {author} {\bibinfo {author} {\bibfnamefont {G.~A.}\ \bibnamefont
  {Rodr{\'{\i}}guez}}, \bibinfo {author} {\bibfnamefont {R.~M.}\ \bibnamefont
  {Hart}}, \bibinfo {author} {\bibfnamefont {A.~J.}\ \bibnamefont {Sievers}},
  \bibinfo {author} {\bibfnamefont {F.}~\bibnamefont {Keilmann}}, \bibinfo
  {author} {\bibfnamefont {Z.}~\bibnamefont {Schlesinger}}, \bibinfo {author}
  {\bibfnamefont {S.~L.}\ \bibnamefont {Wright}},\ and\ \bibinfo {author}
  {\bibfnamefont {W.~I.}\ \bibnamefont {Wang}},\ }\bibfield  {title} {\bibinfo
  {title} {Intensity-dependent cyclotron resonance in a {GaAs}/{GaAlAs}
  two-dimensional electron gas},\ }\href {https://doi.org/10.1063/1.97115}
  {\bibfield  {journal} {\bibinfo  {journal} {Applied Physics Letters}\
  }\textbf {\bibinfo {volume} {49}},\ \bibinfo {pages} {458} (\bibinfo {year}
  {1986})}\BibitemShut {NoStop}%
\bibitem [{\citenamefont {Mics}\ \emph {et~al.}(2015)\citenamefont {Mics},
  \citenamefont {Tielrooij}, \citenamefont {Parvez}, \citenamefont {Jensen},
  \citenamefont {Ivanov}, \citenamefont {Feng}, \citenamefont {M{\"u}llen},
  \citenamefont {Bonn},\ and\ \citenamefont {Turchinovich}}]{Mics2015}%
  \BibitemOpen
  \bibfield  {author} {\bibinfo {author} {\bibfnamefont {Z.}~\bibnamefont
  {Mics}}, \bibinfo {author} {\bibfnamefont {K.-J.}\ \bibnamefont {Tielrooij}},
  \bibinfo {author} {\bibfnamefont {K.}~\bibnamefont {Parvez}}, \bibinfo
  {author} {\bibfnamefont {S.~A.}\ \bibnamefont {Jensen}}, \bibinfo {author}
  {\bibfnamefont {I.}~\bibnamefont {Ivanov}}, \bibinfo {author} {\bibfnamefont
  {X.}~\bibnamefont {Feng}}, \bibinfo {author} {\bibfnamefont {K.}~\bibnamefont
  {M{\"u}llen}}, \bibinfo {author} {\bibfnamefont {M.}~\bibnamefont {Bonn}},\
  and\ \bibinfo {author} {\bibfnamefont {D.}~\bibnamefont {Turchinovich}},\
  }\bibfield  {title} {\bibinfo {title} {Thermodynamic picture of ultrafast
  charge transport in graphene},\ }\href {https://doi.org/10.1038/ncomms8655}
  {\bibfield  {journal} {\bibinfo  {journal} {Nat. Commun.}\ }\textbf {\bibinfo
  {volume} {6}},\ \bibinfo {pages} {7655} (\bibinfo {year} {2015})}\BibitemShut
  {NoStop}%
\bibitem [{\citenamefont {Beregulin}\ \emph {et~al.}(1988)\citenamefont
  {Beregulin}, \citenamefont {Ganichev}, \citenamefont {Glukh}, \citenamefont
  {Gusev}, \citenamefont {Kvon}, \citenamefont {Shik},\ and\ \citenamefont
  {Yaroshetskii}}]{Beregulin1988}%
  \BibitemOpen
  \bibfield  {author} {\bibinfo {author} {\bibfnamefont {E.~V.}\ \bibnamefont
  {Beregulin}}, \bibinfo {author} {\bibfnamefont {S.~D.}\ \bibnamefont
  {Ganichev}}, \bibinfo {author} {\bibfnamefont {K.~Y.}\ \bibnamefont {Glukh}},
  \bibinfo {author} {\bibfnamefont {G.~M.}\ \bibnamefont {Gusev}}, \bibinfo
  {author} {\bibfnamefont {Z.~D.}\ \bibnamefont {Kvon}}, \bibinfo {author}
  {\bibfnamefont {A.~Y.}\ \bibnamefont {Shik}},\ and\ \bibinfo {author}
  {\bibfnamefont {I.~D.}\ \bibnamefont {Yaroshetskii}},\ }\bibfield  {title}
  {\bibinfo {title} {Submillimeter photoconductivity in inversion layers at a
  silicon surface},\ }\href@noop {} {\bibfield  {journal} {\bibinfo  {journal}
  {JETP}\ }\textbf {\bibinfo {volume} {48}},\ \bibinfo {pages} {247} (\bibinfo
  {year} {1988})}\BibitemShut {NoStop}%
\bibitem [{\citenamefont {Beregulin}\ \emph {et~al.}(1990)\citenamefont
  {Beregulin}, \citenamefont {Ganichev}, \citenamefont {Glukh}, \citenamefont
  {Gusev}, \citenamefont {Kvon}, \citenamefont {Shik},\ and\ \citenamefont
  {Yaroshetskii}}]{Beregulin1990}%
  \BibitemOpen
  \bibfield  {author} {\bibinfo {author} {\bibfnamefont {E.~V.}\ \bibnamefont
  {Beregulin}}, \bibinfo {author} {\bibfnamefont {S.~D.}\ \bibnamefont
  {Ganichev}}, \bibinfo {author} {\bibfnamefont {K.~Y.}\ \bibnamefont {Glukh}},
  \bibinfo {author} {\bibfnamefont {G.~M.}\ \bibnamefont {Gusev}}, \bibinfo
  {author} {\bibfnamefont {Z.~D.}\ \bibnamefont {Kvon}}, \bibinfo {author}
  {\bibfnamefont {A.~Y.}\ \bibnamefont {Shik}},\ and\ \bibinfo {author}
  {\bibfnamefont {I.~D.}\ \bibnamefont {Yaroshetskii}},\ }\bibfield  {title}
  {\bibinfo {title} {Rapid submillimeter photoconductivity and energy
  relaxation of a two-dimensional electron gas near the surface of silicon},\
  }\href@noop {} {\bibfield  {journal} {\bibinfo  {journal} {JETP}\ }\textbf
  {\bibinfo {volume} {70}},\ \bibinfo {pages} {1138} (\bibinfo {year}
  {1990})}\BibitemShut {NoStop}%
\bibitem [{Note1()}]{Note1}%
  \BibitemOpen
  \bibinfo {note} {Apart from saturation of the radiation absorption ${\protect
  \cal A}(I)$ mentioned above, these can result from details of the temperature
  dependence of resistance $R(T_\protect \text {e},T)$, which can be governed
  by different microscopic mechanisms in different temperature regimes. In
  other words, at high $I$ Eq.~\protect \textup {\hbox {\mathsurround \z@
  \protect \normalfont (\ignorespaces \ref {dRlinear}\unskip \@@italiccorr )}}
  is not always applicable, and should be replaced by a more general expression
  for the photoresistance, $\Delta R= R[T_\protect \text {e}(I),T]-R[T_\protect
  \text {e}(I\to 0),T]$.}\BibitemShut {Stop}%
\bibitem [{\citenamefont {Baskin}\ \emph {et~al.}(1978)\citenamefont {Baskin},
  \citenamefont {Magarill},\ and\ \citenamefont {Entin}}]{Baskin1978}%
  \BibitemOpen
  \bibfield  {author} {\bibinfo {author} {\bibfnamefont {E.~M.}\ \bibnamefont
  {Baskin}}, \bibinfo {author} {\bibfnamefont {L.~N.}\ \bibnamefont
  {Magarill}},\ and\ \bibinfo {author} {\bibfnamefont {M.~V.}\ \bibnamefont
  {Entin}},\ }\bibfield  {title} {\bibinfo {title} {Two-dimensional
  electron-impurity system in a strongmagnetic field},\ }\href@noop {}
  {\bibfield  {journal} {\bibinfo  {journal} {Sov. Phys. JETP}\ }\textbf
  {\bibinfo {volume} {48}},\ \bibinfo {pages} {365} (\bibinfo {year}
  {1978})}\BibitemShut {NoStop}%
\bibitem [{\citenamefont {Bobylev}\ \emph {et~al.}(1995)\citenamefont
  {Bobylev}, \citenamefont {Maa{\o}}, \citenamefont {Hansen},\ and\
  \citenamefont {Hauge}}]{Bobylev1995}%
  \BibitemOpen
  \bibfield  {author} {\bibinfo {author} {\bibfnamefont {A.~V.}\ \bibnamefont
  {Bobylev}}, \bibinfo {author} {\bibfnamefont {F.~A.}\ \bibnamefont
  {Maa{\o}}}, \bibinfo {author} {\bibfnamefont {A.}~\bibnamefont {Hansen}},\
  and\ \bibinfo {author} {\bibfnamefont {E.~H.}\ \bibnamefont {Hauge}},\
  }\bibfield  {title} {\bibinfo {title} {Two-dimensional magnetotransport
  according to the classical lorentz model},\ }\href
  {https://doi.org/10.1103/physrevlett.75.197} {\bibfield  {journal} {\bibinfo
  {journal} {Phys. Rev. Lett.}\ }\textbf {\bibinfo {volume} {75}},\ \bibinfo
  {pages} {197} (\bibinfo {year} {1995})}\BibitemShut {NoStop}%
\bibitem [{\citenamefont {Mirlin}\ \emph {et~al.}(2001)\citenamefont {Mirlin},
  \citenamefont {Polyakov}, \citenamefont {Evers},\ and\ \citenamefont
  {W\"olfle}}]{Mirlin2001}%
  \BibitemOpen
  \bibfield  {author} {\bibinfo {author} {\bibfnamefont {A.~D.}\ \bibnamefont
  {Mirlin}}, \bibinfo {author} {\bibfnamefont {D.~G.}\ \bibnamefont
  {Polyakov}}, \bibinfo {author} {\bibfnamefont {F.}~\bibnamefont {Evers}},\
  and\ \bibinfo {author} {\bibfnamefont {P.}~\bibnamefont {W\"olfle}},\
  }\bibfield  {title} {\bibinfo {title} {Quasiclassical negative
  magnetoresistance of a 2d electron gas: Interplay of strong scatterers and
  smooth disorder},\ }\href {http://dx.doi.org/10.1103/PhysRevLett.87.126805}
  {\bibfield  {journal} {\bibinfo  {journal} {Phys. Rev. Lett.}\ }\textbf
  {\bibinfo {volume} {87}},\ \bibinfo {pages} {126805} (\bibinfo {year}
  {2001})}\BibitemShut {NoStop}%
\bibitem [{\citenamefont {Dmitriev}\ \emph {et~al.}(2004)\citenamefont
  {Dmitriev}, \citenamefont {Mirlin},\ and\ \citenamefont
  {Polyakov}}]{Dmitriev2004}%
  \BibitemOpen
  \bibfield  {author} {\bibinfo {author} {\bibfnamefont {I.~A.}\ \bibnamefont
  {Dmitriev}}, \bibinfo {author} {\bibfnamefont {A.~D.}\ \bibnamefont
  {Mirlin}},\ and\ \bibinfo {author} {\bibfnamefont {D.~G.}\ \bibnamefont
  {Polyakov}},\ }\bibfield  {title} {\bibinfo {title} {Oscillatory ac
  conductivity and photoconductivity of a two-dimensional electron gas:
  Quasiclassical transport beyond the boltzmann equation},\ }\href
  {http://dx.doi.org/10.1103/PhysRevB.70.165305} {\bibfield  {journal}
  {\bibinfo  {journal} {Phys. Rev. B}\ }\textbf {\bibinfo {volume} {70}},\
  \bibinfo {pages} {165305} (\bibinfo {year} {2004})}\BibitemShut {NoStop}%
\bibitem [{\citenamefont {Beltukov}\ and\ \citenamefont
  {Dyakonov}(2016)}]{Beltukov2016}%
  \BibitemOpen
  \bibfield  {author} {\bibinfo {author} {\bibfnamefont {Y.~M.}\ \bibnamefont
  {Beltukov}}\ and\ \bibinfo {author} {\bibfnamefont {M.~I.}\ \bibnamefont
  {Dyakonov}},\ }\bibfield  {title} {\bibinfo {title} {Microwave-induced
  resistance oscillations as a classical memory effect},\ }\href
  {https://doi.org/10.1103/physrevlett.116.176801} {\bibfield  {journal}
  {\bibinfo  {journal} {Phys. Rev. Lett.}\ }\textbf {\bibinfo {volume} {116}},\
  \bibinfo {pages} {176801} (\bibinfo {year} {2016})}\BibitemShut {NoStop}%
\bibitem [{\citenamefont {Dorozhkin}\ \emph {et~al.}(2017)\citenamefont
  {Dorozhkin}, \citenamefont {Kapustin}, \citenamefont {Dmitriev},
  \citenamefont {Umansky}, \citenamefont {von Klitzing},\ and\ \citenamefont
  {Smet}}]{Dorozhkin2017}%
  \BibitemOpen
  \bibfield  {author} {\bibinfo {author} {\bibfnamefont {S.~I.}\ \bibnamefont
  {Dorozhkin}}, \bibinfo {author} {\bibfnamefont {A.~A.}\ \bibnamefont
  {Kapustin}}, \bibinfo {author} {\bibfnamefont {I.~A.}\ \bibnamefont
  {Dmitriev}}, \bibinfo {author} {\bibfnamefont {V.}~\bibnamefont {Umansky}},
  \bibinfo {author} {\bibfnamefont {K.}~\bibnamefont {von Klitzing}},\ and\
  \bibinfo {author} {\bibfnamefont {J.~H.}\ \bibnamefont {Smet}},\ }\bibfield
  {title} {\bibinfo {title} {Evidence for non-markovian electron dynamics in
  the microwave absorption of a two-dimensional electron system},\ }\href
  {https://link.aps.org/doi/10.1103/PhysRevB.96.155306} {\bibfield  {journal}
  {\bibinfo  {journal} {Phys. Rev. B}\ }\textbf {\bibinfo {volume} {96}},\
  \bibinfo {pages} {155306} (\bibinfo {year} {2017})}\BibitemShut {NoStop}%
\bibitem [{\citenamefont {Chepelianskii}\ and\ \citenamefont
  {Shepelyansky}(2018)}]{Chepelianskii2018}%
  \BibitemOpen
  \bibfield  {author} {\bibinfo {author} {\bibfnamefont {A.~D.}\ \bibnamefont
  {Chepelianskii}}\ and\ \bibinfo {author} {\bibfnamefont {D.~L.}\ \bibnamefont
  {Shepelyansky}},\ }\bibfield  {title} {\bibinfo {title} {Floquet theory of
  microwave absorption by an impurity in the two-dimensional electron gas},\
  }\href {https://link.aps.org/doi/10.1103/PhysRevB.97.125415} {\bibfield
  {journal} {\bibinfo  {journal} {Phys. Rev. B}\ }\textbf {\bibinfo {volume}
  {97}},\ \bibinfo {pages} {125415} (\bibinfo {year} {2018})}\BibitemShut
  {NoStop}%
\bibitem [{\citenamefont {Zayats}\ \emph {et~al.}(2009)\citenamefont {Zayats},
  \citenamefont {Richards},\ and\ \citenamefont {Richards}}]{Zayats2009}%
  \BibitemOpen
  \bibfield  {author} {\bibinfo {author} {\bibfnamefont {A.}~\bibnamefont
  {Zayats}}, \bibinfo {author} {\bibfnamefont {D.}~\bibnamefont {Richards}},\
  and\ \bibinfo {author} {\bibfnamefont {D.}~\bibnamefont {Richards}},\ }\href
  {https://books.google.de/books?id=DECRmAEACAAJ} {\emph {\bibinfo {title}
  {Nano-optics and Near-field Optical Microscopy}}},\ Artech House nanoscale
  science and engineering\ (\bibinfo  {publisher} {Artech House},\ \bibinfo
  {year} {2009})\BibitemShut {NoStop}%
\bibitem [{\citenamefont {Keller}(2012)}]{Keller2012}%
  \BibitemOpen
  \bibfield  {author} {\bibinfo {author} {\bibfnamefont {O.}~\bibnamefont
  {Keller}},\ }\href {https://books.google.de/books?id=v2ck__wFOBEC} {\emph
  {\bibinfo {title} {Quantum Theory of Near-Field Electrodynamics}}},\
  Nano-Optics and Nanophotonics\ (\bibinfo  {publisher} {Springer Berlin
  Heidelberg},\ \bibinfo {year} {2012})\BibitemShut {NoStop}%
\bibitem [{\citenamefont {Girard}\ and\ \citenamefont
  {Dereux}(1996)}]{Girard1996}%
  \BibitemOpen
  \bibfield  {author} {\bibinfo {author} {\bibfnamefont {C.}~\bibnamefont
  {Girard}}\ and\ \bibinfo {author} {\bibfnamefont {A.}~\bibnamefont
  {Dereux}},\ }\bibfield  {title} {\bibinfo {title} {Near-field optics
  theories},\ }\href {https://doi.org/10.1088/0034-4885/59/5/002} {\bibfield
  {journal} {\bibinfo  {journal} {Reports on Progress in Physics}\ }\textbf
  {\bibinfo {volume} {59}},\ \bibinfo {pages} {657} (\bibinfo {year}
  {1996})}\BibitemShut {NoStop}%
\end{thebibliography}%

	
\end{document}